\documentclass[preprint]{aastex631}

\usepackage{comment}
\usepackage{rotating}
\usepackage{ulem}
\usepackage{soul}
\usepackage{multirow}

\usepackage{longtable}
\usepackage{indentfirst}  
\usepackage{longtable}  
\shorttitle{Eight New Pulsars in NGC 6517}
\shortauthors{Yin et al.}

\graphicspath{{./}{figures/}}

\begin{document}
\title{FAST Discovery of Eight Isolated Millisecond Pulsars in NGC 6517}

\author[0000-0001-6051-3420]{Dejiang Yin}
\affiliation{College of Physics, Guizhou University, Guiyang 550025, China}

\author[0000-0002-2394-9521]{Li-yun Zhang}
\affiliation{College of Physics, Guizhou University, Guiyang 550025, China}
\affiliation{International Centre of Supernovae, Yunnan Key Laboratory, Kunming 650216, China}

\author[0000-0003-0597-0957]{Lei Qian}
\affiliation{Guizhou Radio Astronomical Observatory, Guizhou University Guiyang 550025, China}
\affiliation{National Astronomical Observatories, Chinese Academy of Sciences, 20A Datun Road, Chaoyang District, Beijing 100101, China}
\affiliation{CAS Key Laboratory of FAST, National Astronomical Observatories, Chinese Academy of Sciences, Beijing 100101, China}
\affiliation{College of Astronomy and Space Sciences, University of Chinese Academy of Sciences, Beijing 100049, China}

\author[0000-0001-6196-4135]{Ralph~P.~Eatough}
\affiliation{National Astronomical Observatories, Chinese Academy of Sciences, 20A Datun Road, Chaoyang District, Beijing 100101, China}
\affiliation{Max-Planck-Institut f\"{u}r Radioastronomie, Auf dem H\"{u}gel 69, D-53121, Bonn, Germany}

\author{Baoda Li}
\affiliation{College of Physics, Guizhou University, Guiyang 550025, China}

\author[0000-0003-1301-966X]{Duncan R. Lorimer}
\affiliation{Department of Physics and Astronomy, West Virginia University, Morgantown, WV 26506-6315, USA}
\affiliation{Center for Gravitational Waves and Cosmology, West Virginia University, Chestnut Ridge Research Building, Morgantown, WV 26505, USA}

\author[0009-0007-6396-7891]{Yinfeng Dai}
\affiliation{College of Physics, Guizhou University, Guiyang 550025, China}

\author{Yaowei Li}
\affiliation{College of Physics, Guizhou University, Guiyang 550025, China}

\author{Xingnan Zhang}
\affiliation{State Key Laboratory of Public Big Data, Guizhou University, Guiyang 550025, China}

\author{Minghui Li}
\affiliation{State Key Laboratory of Public Big Data, Guizhou University, Guiyang 550025, China}

\author{Tianhao Su}
\affiliation{College of Physics, Guizhou University, Guiyang 550025, China}

\author{Yuxiao Wu}
\affiliation{Chongqing University of Posts and Telecommunications, Chongqing, 40000, China}

\author{Yu Pan}
\affiliation{Chongqing University of Posts and Telecommunications, Chongqing, 40000, China}

\author[0009-0001-6693-7555]{Yujie Lian} 
\affiliation{Institute for Frontiers in Astronomy and Astrophysics, Beijing Normal University, Beijing 102206, China}
\affiliation{Department of Astronomy, Beijing Normal University, Beijing 100875, China}

\author{Tong Liu}
\affiliation{National Astronomical Observatories, Chinese Academy of Sciences, 20A Datun Road, Chaoyang District, Beijing 100101, China}

\author[0000-0002-9322-9319]{Zhen Yan}
\affiliation{Shanghai Astronomical Observatory, Chinese Academy of Sciences, Shanghai 200030, China}

\author[0000-0001-7771-2864]{Zhichen Pan}
\affiliation{Guizhou Radio Astronomical Observatory, Guizhou University Guiyang 550025, China}
\affiliation{National Astronomical Observatories, Chinese Academy of Sciences, 20A Datun Road, Chaoyang District, Beijing 100101, China}
\affiliation{CAS Key Laboratory of FAST, National Astronomical Observatories, Chinese Academy of Sciences, Beijing 100101, China}
\affiliation{College of Astronomy and Space Sciences, University of Chinese Academy of Sciences, Beijing 100049, China}
\correspondingauthor{}\email{liy\_zhang@hotmail.com; panzc@bao.ac.cn; lqian@nao.cas.cn}

\begin{abstract}
  We present the discovery of 8 isolated millisecond pulsars in Globular Cluster (GC) NGC 6517 using the Five-Hundred-meter Aperture Spherical radio Telescope (FAST). 
  The spin periods of those pulsars (namely PSR~J1801--0857K to R, or, NGC~6517K to R) are all shorter than 10\,ms.
  With these discoveries, NGC~6517 is currently the GC with the most known pulsars in the FAST sky.
  The largest difference in dispersion measure of the pulsars in NGC~6517 is 11.2\,cm$^{-3}$ pc, the second among all GCs. 
 The fraction of isolated pulsars in this GC (16 of 17, 94$\%$) is consistent with previous studies indicating an overabundance of isolated pulsars in the densest GCs, especially in those undergoing cluster core collapse.
  Considering the FAST GC pulsar discoveries,
  we modeled the GC pulsar population using the empirical Bayesian method described by
  Turk and Lorimer with the recent counts.
  Using this approach, we find that the expected number of potential pulsars in GCs seems to be correlated with the central escape velocity, hence, the GCs Liller 1, NGC 6441, M54 (NGC 6715), and $\omega$-Cen (NGC 5139) are expected to host the largest numbers of pulsars. 
  
\end{abstract}

\keywords{Globular star clusters (656); Millisecond pulsars (1062); Radio telescopes (1360)}
\section{Introduction}

The proportion of millisecond pulsars in Globular Clusters (GCs) with spin periods shorter than 30\,ms is $\sim 97\%$ (e.g., \citealt{2023RAA....23e5012Y}), much higher than that in the field population ($\sim 12\%$\footnote{\citealt{2005AJ....129.1993M}, 1.70 version, \url{https://www.atnf.csiro.au/people/pulsar/psrcat/}}),
making them rich targets for fast spinning pulsars and exotic binary systems, such as Terzan~5ad (with the highest known spin frequency of 716~Hz, \citealt{2007ApJ...670..363H}); 
M4A (the triple system with a planet, \citealt{2003Sci...301..193S}); 
M71E (a binary in M71 with an orbit of 53 minutes, \citealt{2023Natur.620..961P});
M15C (a double neutron star binary, \citealt{1990Natur.346...42A});
and NGC~1851E (a pulsar--black hole system candidate, \citealt{2024Sci...383..275B}).
GCs have been the subject of pulsar searches for more than 30 years. 
Since the first GC pulsar was discovered \citep{1987Natur.328..399L}, 
322 radio pulsars have been detected in 41 GCs as of May 2023\footnote{\url{https://www3.mpifr-bonn.mpg.de/staff/pfreire/GCpsr.html}}. 
Previous surveys for GC pulsars have been carried out 
with telescopes across the world, 
including Lovell (e.g., \citealt{1987Natur.328..399L}), 
Parkes (e.g., \citealt{1995MNRAS.274..547R}), 
Arecibo (e.g., \citealt{2007ApJ...670..363H}), 
Green Bank Telescope (GBT, e.g., \citealt{2005Sci...307..892R}), 
and Giant Metrewave Radio Telescope (GMRT, e.g., \citealt{2022A&A...664A..54G}).

In the past 5 years alone, about 120 GC pulsars have been discovered with newly built telescopes.
MeerKAT (\citealt{2016mks..confE...1J}) found $\sim$ 80 GC pulsars (e.g., \citealt{2021MNRAS.504.1407R}, TRAPUM project\footnote{\url{http://trapum.org/discoveries/}}) and the telescope used in this work, Five-hundred-meter Aperture Spherical radio Telescope (FAST, \citealt{2011IJMPD..20..989N}) discovered another $\sim$ 40 (e.g., \citealt{2020ApJ...892L...6P}, FAST GC Pulsar Discoveries\footnote{\url{https://fast.bao.ac.cn/cms/article/65/}}). With a sensitivity limit of 4 $\mu$Jy, 100 to 300 GC pulsars are expected to be discovered by SKA1-MID when it comes online \citep{2015aska.confE..47H}.

Among 170 Galactic GCs, there are 45 in the FAST sky. One of these 45 GCs,
NGC 6517 is located at $\alpha_{\rm J2000} = 18^{\rm h}01^{\rm m}50.52^{\rm s}$, 
$\delta_{\rm J2000} = -08^{\rm \circ}57^{\rm '}31.6^{\rm ''}$. It is the densest GC ($\rho_0 = 10^{5.29}L_{\odot} {\rm pc}^{-3}$) in the FAST sky.
Its core and half-light radius are $r_{\rm c} = 0.06{\rm'}$ and $r_{\rm h} = 0.50{\rm'}$,  
respectively (\citealt{1996AJ....112.1487H}, 2010 edition), well within the FAST half-power beamwidth at L-band ($\sim$2.9\arcmin, \citealt{2020Innov...100053Q}).
The four pulsars, PSR~J1801$-$0857A to D (NGC~6517A to D), were detected by the GBT \citep{2011ApJ...734...89L}.
The dispersion measures (DMs) of those pulsars range from 174.71 to 182.56\ cm$^{-3}$ pc.
PSR~J1801$-$0857E, F, and G (NGC~6517E, F, and G) were discovered by FAST with a candidate selection method based on DM--signal-to-noise ratio curves \citep{2021RAA....21..143P}.
PSR~J1801$-$0857H and I (NGC 6517H and I) were also discovered at FAST in long tracking observations \citep{2021ApJ...915L..28P}.
Currently, NGC~6517B is the only  binary pulsar ($\sim$60\,d orbital period) detected in this GC. Among all 41 GCs currently
known to host pulsars, NGC~6517 ranks third highest in terms of the fraction of isolated pulsars, at 94\%. 
Only NGC~6522, where 6 out of the 6 known pulsars are isolated, and Terzan~1 (8 out of 8) have a higher fraction.
This is in spite of the fact that, as described below, our analyses of NGC~6517 include acceleration searches which aim to retain sensitivity to binary systems.

Located at 10.4 and 10.6\ kpc away from the Sun (\citealt{1996AJ....112.1487H}, 2010 edition),
NGC~6517 and M15 (NGC~7078) respectively, are the only two core-collapsed GCs (e.g., \citealt{2014A&A...561A..11V}) in the FAST sky. 
The isolated pulsars in NGC~6517 and M15 dominate the pulsar population, 
similar to other core-collapsed GCs outside the FAST sky, 
such as Terzan 1 (8 isolated, 0 binaries) and NGC 6624 (10 isolated, 2 binaries).
In \cite{2014A&A...561A..11V} a dynamical parameter, the encounter rate per binary in a GC, $\gamma$ ($\gamma \propto \rho_0^{0.5} r_c^{-1}$, $\rho_0$ is the central mass density and $r_c$ is the core radius), was used as 
a factor to characterize  the differences between the pulsar populations of different GCs. 
Original millisecond pulsar binaries are thought to be disrupted by an exchange encounter, enabled by the higher $\gamma$ in core-collapsed clusters.
The high fraction of isolated pulsars in core-collapsed GCs is
therefore a natural result of the increased number of these encounter and disruption events.
In addition, several alternative formation scenarios are proposed to explain the seeming overabundance of isolated millisecond pulsars in core-collapsed clusters, such as the formation of single millisecond pulsars via accretion following mergers of neutron stars and main-sequence stars (e.g., \citealt{1992ApJ...401..246D, 2005ASPC..328..147C, 2022ApJ...934L...1K}), the formation of single pulsars via mergers of massive white dwarf binaries (e.g., \citealt{2001MNRAS.320L..45K, 2021ApJ...906...53S, 2023MNRAS.525L..22K}) 
, and the number of single millisecond pulsars in GCs can significantly be boosted by both main-sequence star tidal disruption events and merger-induced collapses discussed in the recent N-body simulations \citep{2024ApJ...961...98Y}.

Ranking third among the GCs with the most pulsars in the FAST sky, NGC~6517 is one
of pulsar search targets with the highest chance for further discoveries \citep[see, e.g.,][]{2013MNRAS.436.3720T}. 
Motivated by this potentially significant pulsar population, in this work we present the discovery of an additional eight pulsars
in NGC~6517, bringing the total number to 17 in this cluster. The new discoveries were made during the reprocessing of FAST archival data 
and increase the total number of GC pulsars discovered by FAST to 49. Pulsar population simulations can help to predict the expected number of GC pulsars. Due to sensitivity limits and the selection effects of binary pulsar search algorithms, the number of known pulsars in a given GC can be much lower than predictions from simulations.
The stellar encounter rate $\Gamma$, which is estimated with $\Gamma \propto \rho_0^{1.5} r_c^2$ ($\rho_0$ is the central mass density and $r_c$ is the core radius, \citealt{2003ApJ...591L.131P}),
has also been used to estimate the GC pulsar numbers (e.g., \citealt{2003ApJ...591L.131P, 2009Sci...325..848A}).
The number of potential pulsars in a GC is scaled by its $\Gamma$ in the empirical Bayesian model of \citet{2013MNRAS.436.3720T}.
This method was used to estimate the total population of pulsars for all Galactic GCs and 
predicted a number of 2280$_{-1490}^{+2720}$ \citep{2013MNRAS.436.3720T}. 
In light of the number of GC pulsars discovered by FAST, it is timely to revisit the population analysis carried out a decade ago by \citet{2013MNRAS.436.3720T}.

The observations and data reduction are presented in Section \ref{sec:obs}.
The discoveries of pulsars and their timing solutions are given in Section \ref{Results}.
Section \ref{psr_numbsers} provides 
an updated analysis of the pulsar population in GCs 
using the procedures described in \citet{2013MNRAS.436.3720T}.
A summary of the findings is given in Section \ref{Conclusions}.

\section{Observations and Data Reduction} \label{sec:obs}
\subsection{Observations}

As a part of the project ``Globular Cluster FAST: A Neutron-star Survey'' (GC FANS, \citealt{2021ApJ...915L..28P, 2023arXiv231206067W}), 
the first observation of NGC~6517 was performed on June 24, 2019.
By December 31, 2022, a total of 19 observations with integration lengths between 1400\,s and 9000\,s were carried out.
The data were taken with the central beam of the FAST 19-beam receiver covering a frequency range of 1.0-1.5\,GHz. 
The data were channelized into 4096 channels (channel width $0.122\,{\rm MHz}$).
All observations were performed with the tracking mode with a sampling time of $49.152\,\mu{\rm s}$ and packaged into the standard search-mode PSRFITS data format \citep{2004PASA...21..302H}.
During each observation, the central beam was pointed at the center of NGC~6517.

\subsection{Data Reduction} \label{sec:red}

\textsc{PRESTO}\footnote{\url{https://github.com/scottransom/presto}}
\citep{2001PhDT.......123R, 2002AJ....124.1788R} and 
\textsc{TEMPO}\footnote{\url{http://nanograv.github.io/tempo/}} \citep{2015ascl.soft09002N}
were used for pulsar search and timing respectively. In the following, we first describe the pulsar search procedure, followed by the timing analysis.  

The routine \texttt{rfifind} from \textsc{PRESTO} was used to identify and mask the detrimental effects of terrestrial radio frequency interference (RFI) in both the time and frequency domains. The \texttt{prepdata} or \texttt{prepsubband} routines were used to dedisperse the PSRFITS data over a range of trial DM values.
Since the DM of previously known pulsars in NGC~6517 ranges from $174.7\,{\rm cm}^{-3}\,{\rm pc}$ (NGC 6517D) to 185.6\,cm$^{-3}$\ pc (NGC 6517G), with an average of $181.3~{\rm cm}^{-3}\,{\rm pc}$,  dedispersion was performed across a DM range of 170 to 190~cm$^{-3}$\,pc, with a step of 0.05\,cm$^{-3}$\,pc.
This DM coverage was determined to maintain sensitivity to any undiscovered pulsars in NGC~6517.

To identify periodic signals, the dedispersed time series were transformed into the fluctuation frequency domain using the routine \texttt{realfft}. 
The routine \texttt{accelsearch} was then used to perform Fourier domain acceleration searches, with harmonic summing to account for pulses of narrow duty cycle. 
The \texttt{accelsearch} parameter \texttt{zmax} represents the largest drift of the pulse frequency, in Fourier bins, under the assumption of a constant acceleration (\citealt{2002AJ....124.1788R}). We performed a search for isolated pulsars with a \texttt{zmax} of 20 and a segmented acceleration search (with time series of length down to $1/5$ of the original integration time which ranged from 500 \,s to 1800\,s) for binaries and with a \texttt{zmax} of 1200. 

So-called candidate ``sifting'' routines were then used to amalgamate multiple detections of each pulsar candidate from the \texttt{accelsearch} results. Both the routine \texttt{ACCEL\_sift.py} of \textsc{PRESTO} and \texttt{JinglePulsar\footnote{\url{https://github.com/jinglepulsar/jinglesifting}}} were used for this task.
We also identified any candidates that appeared on different days, but with similar spin period and DM values. This method will be described in another publication (Yu et al., in preparation).

In the final stage of the pulsar search, the \texttt{prepfold} routine was used to fold all sifted pulsar candidates using either dedispersed time series, or raw PSRFITs data, to generate standard pulsar candidate diagnostic plots for visual inspection.

The timing analysis of detected pulsars was performed as follows: 
\begin{enumerate}
\item  The observational data were dedispersed to time series according to the candidate DM value.
During this process, the RFI ``mask'' generated earlier from the routine \texttt{rfifind} was used.
\item The dedispersed and RFI cleaned time series were folded at the pulsar spin period in the topocentric reference frame by the routine \texttt{prepfold}.

\item The \textsc{PRESTO} routines \texttt{pygaussfit.py} and \texttt{get\_TOAs.py} were utilized to form a standard reference template profile and extract the times of arrival (TOAs), respectively. 

\item The initial pulsar ephemeris was then iterated using \textsc{TEMPO} over progressively longer time spans
until a phase-connected timing solution was obtained. 
\end{enumerate}
A phase-connected timing solution means that every rotation of the pulsar can be accurately predicted by an ephemeris.
The \texttt{pyplotres.py} routine from \textsc{PRESTO} was used to realize interactive inspection of timing residuals. 
The ephemerides of pulsars were gradually fitted and updated by removing the arbitrary phase offsets (with so-called ``JUMPs'') between epochs.
In the archival observations of NGC~6517, the gap between observations could be up to more than one year. The script \texttt{Dracula}\footnote{\url{https://github.com/pfreire163/Dracula}} was also used to determine the exact rotation count of pulsars \citep{2018MNRAS.476.4794F}.

\section{Results}
\label{Results}

\begin{figure}[ht]
\centering
\includegraphics[width=18cm, angle=0]{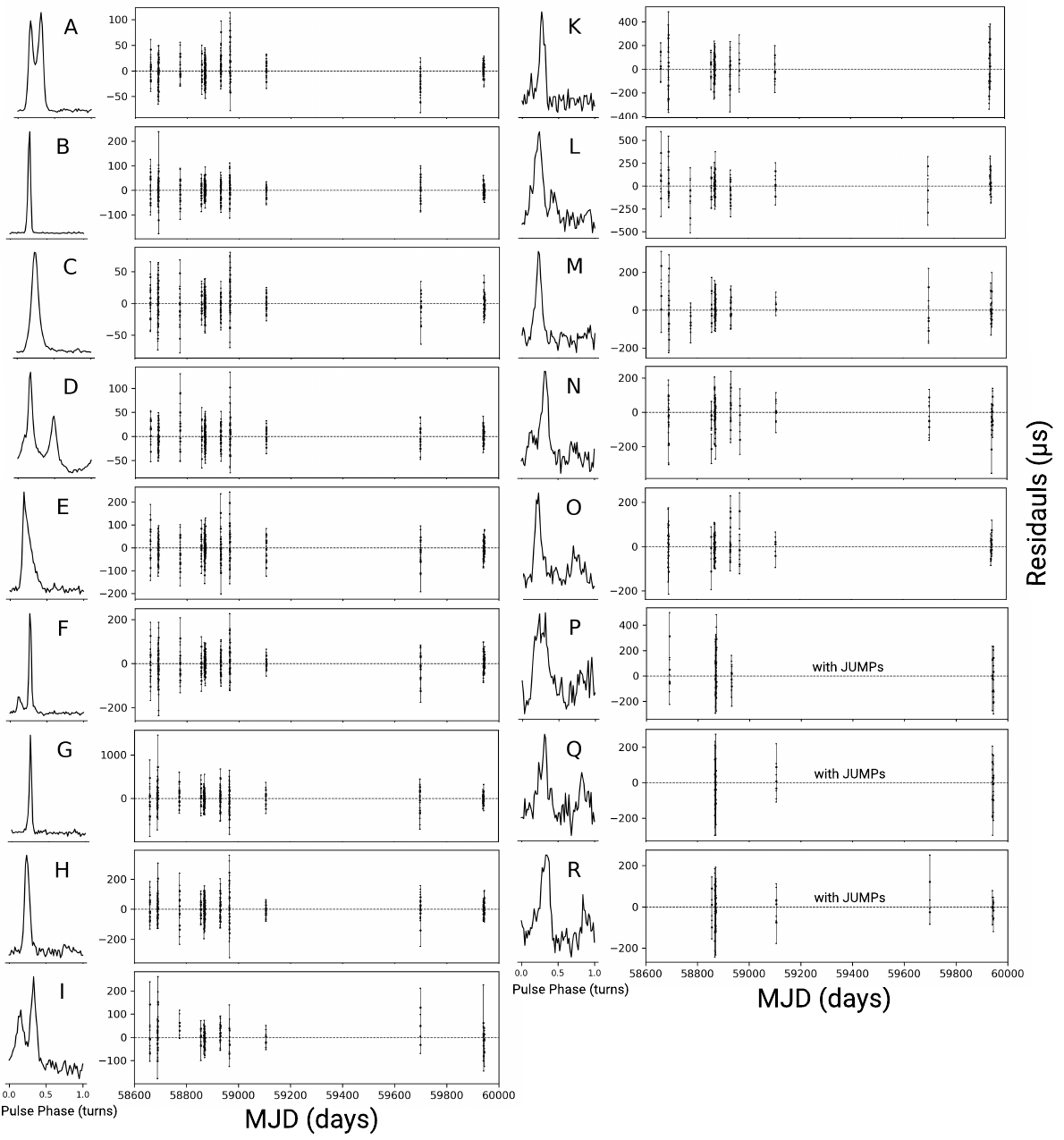}
\caption{The average pulse profiles and timing residuals of all pulsars in NGC~6517. 
The left panel of each subplot is the integrated pulse profile obtained by summing all detections over 64 pulse phase bins. 
The timing residuals from the best-fit timing model are shown in the right panel. 
The phase-connected timing solutions of NGC~6517A to O were obtained, 
while for NGC~6517P, Q, and R, solutions with only the spin frequencies ($f$) were fitted, with commonly used arbitrary phase offsets (so-called ``JUMPs'') between observations.
\label{fig:1}}
\end{figure}

Eight isolated millisecond pulsars in NGC~6517, namely PSR~J1801--0857K to R (NGC~6517K to R)\footnote{The previously announced pulsar NGC~6517J, was subsequently identified as the 17$^{\rm th}$ spin frequency harmonic of NGC~6517B.}, 
were discovered during the reprocessing of FAST archival data. 
The same data set was used to time these newly discovered pulsars and the other known pulsars in NGC~6517. 
Our discoveries make NGC~6517 the GC with the most known pulsars in the FAST sky, hosting a total of 17 pulsars.
The spin periods of the new pulsars are all shorter than 10\,ms.
Their pulse profiles and timing residuals are in Figure~\ref{fig:1}.
The timing solution of NGC~6517B, the only known binary in the GC, was obtained with the ``BT'' model~\citep{1976ApJ...205..580B}, 
which includes the projected pulsar semi-major axis ($x_{\rm p}$), eccentricity ($e$), epoch of periastron passage ($T_{\rm 0}$), orbital period ($P_{\rm b}$) and longitude of periastron passage ($\omega$).

\begin{longtable}{l c c c}
\caption{The timing solutions of known pulsars NGC~6517A to I and new pulsars NGC~6517K to O.
In the timing analysis, the DE438 Solar System Ephemeris and the TBD time units are used and the times were not referenced to one of the atomic time standards (UNCORR).
The timing data spans from MJD 58659 to 59944 for NGC~6517A to M, MJD 58688 to 59944 for NGC~6517N and NGC~6517O.
\label{tab:1}} \\
\hline
\endfirsthead %
\multicolumn{4}{c}{} \\ %
\multicolumn{4}{c}{{\tablename\ \thetable{} -- Continued from previous page}} \\
\hline
\endhead %

\hline
\multicolumn{4}{r}{{Continued on next page}} \\
\endfoot %

\hline
\endlastfoot %
\hline
\textbf{Pulsar}  &   \textbf{J1801--0857A}  &   \textbf{J1801--0857B} &   \textbf{J1801--0857C}  \\
\hline\hline
Right ascension, $\alpha$ (J2000)                \dotfill &  18:01:50.61097(7) &   18:01:50.5654(1) &   18:01:50.73891(5) \\
Declination, $\delta$ (J2000)                    \dotfill &  --08:57:31.905(3)  &   --08:57:32.867(7)&   --08:57:32.761(3)   \\
Spin frequency, $f$ (s$^{-1}$)                   \dotfill &  139.36088855893(2) &   34.52849284047(2) &   267.47267670678(3)  \\
1st spin frequency derivative, $\dot{f}$ (Hz s$^{-2}$) \dotfill &  9.9034(4)$\times 10^{-15}$   &   $-$2.6270(2)$\times 10^{-15}$ &   4.5058(8)$\times 10^{-15}$  \\
Reference epoch (MJD)                            \dotfill &   58871.058570      &   58871.058570   &   58871.058570     \\
Dispersion measure, DM (pc cm$^{-3}$)             \dotfill &   182.655(3)      &   182.402(3)      &   182.356(2)    \\
Number of TOAs                                    \dotfill &   149      &   152              &   150      \\
Post-fit residual RMS ($\mu$s)                            \dotfill &   17.78        &   25.10          &   15.15    \\
\hline
\multicolumn{4}{c}{Binary Parameters}  \\
\hline\hline
Binary model                                      \dotfill &   --   &   BT   &   --       \\
Projected semi-major axis, $x_{\rm p}$ (lt-s)     \dotfill &   --   &   33.87542(1)    &   --    \\
Orbital eccentricity, $e$                          \dotfill &   --    &   3.82259(6)$\times 10^{-2}$   &   --     \\
Longitude of periastron, $\omega$ (deg)            \dotfill &   --   &   $-$57.8927(7)  &   --  \\
Epoch of periastron passage, $T_0$ (MJD)        \dotfill &   --   &   54757.7229(1)   &   --    \\
Orbital period, $P_b$ (days)                       \dotfill &   --   &   59.8364531(8)     &   --      \\
\hline
\multicolumn{4}{c}{Derived Parameters}  \\
\hline\hline
Spin period, $P$ (ms)   \dotfill &   7.175614409039(1)   &   28.96158846609(1)   &   3.7386996395757(5)    \\
1st Spin period derivative, $\dot{P}$ (s s$^{-1}$)  \dotfill & $-$5.0992(2)$\times 10^{-19}$ &   2.2034(2)$\times 10^{-18}$  & $-$6.298(1)$\times 10^{-20}$ \\
\hline
  &        &       &     \\
\hline
\textbf{Pulsar}  &   \textbf{J1801--0857D}      &   \textbf{J1801--0857E}    &   \textbf{J1801--0857F}    \\
\hline\hline
Right ascension, $\alpha$ (J2000)       \dotfill &  18:01:55.36430(7)    &   18:01:50.6240(1)  &   18:01:50.7417(1) \\
Declination, $\delta$ (J2000)           \dotfill &   --08:57:24.316(3)    &   --08:57:31.331(8) &  --08:57:31.292(8)   \\
Spin frequency, $f$ (s$^{-1}$)          \dotfill &   236.60059797388(3)  &   131.55003246018(5) &   40.17357389120(1)  \\
1st Spin frequency derivative, $\dot{f}$ (Hz s$^{-2}$) \dotfill & $-$4.211(8)$\times 10^{-16}$  &  1.7928(1)$\times 10^{-14}$  &  4.3105(3)$\times 10^{-15}$ \\
Reference Epoch (MJD)                   \dotfill &   58871.058570        &   58871.058570      &   58871.058570     \\
Dispersion measure, DM (pc cm$^{-3}$)   \dotfill &    174.537(3)         &   183.184(5)        &    183.785(5)  \\
Number of TOAs                          \dotfill &   151                 &   146               &   152           \\
Post-fit residual RMS ($\mu$s)                  \dotfill &   17.36               &   40.32             &   34.35       \\
\hline
\multicolumn{4}{c}{Derived Parameters}  \\
\hline\hline
Spin period, $P$ (ms)   \dotfill & 4.2265320061042(6) & 7.601670492196(3) & 24.891985032456(8) \\
1st Spin period derivative, $\dot{P}$ (s s$^{-1}$) \dotfill &   7.52(1)$\times 10^{-21}$   &   $-$1.03598(6)$\times 10^{-18}$  &  $-$2.6708(2)$\times 10^{-18}$  \\
\hline
  &        &       &     \\
\hline
\textbf{Pulsar}  &   \textbf{J1801--0857G}   &   \textbf{J1801--0857H}   &   \textbf{J1801--0857I}    \\
\hline\hline
Right ascension, $\alpha$ (J2000)            \dotfill & 18:01:50.0979(7)   &  18:01:52.5989(2)    &    18:01:53.5073(2)  \\
Declination, $\delta$ (J2000)                \dotfill &  --08:57:27.53(3)  &   --08:57:45.04(1)     &    --08:57:41.84(1)   \\
Spin frequency, $f$ (s$^{-1}$)               \dotfill &   19.38308799410(3)    &   177.21938880576(8)   &   307.2974449774(2)   \\
1st Spin frequency derivative, $\dot{f}$ (Hz s$^{-2}$) \dotfill &   $-$4.01(7)$\times 10^{-17}$ &   $-$1.078(2)$\times 10^{-15}$  &  1.550(4)$\times 10^{-15}$ \\
Reference epoch (MJD)                         \dotfill &   58871.058570     &   58871.058570     &   58871.058570    \\
Dispersion measure, DM (pc cm$^{-3}$)         \dotfill &  185.06(2)  &   179.634(9)   &   177.877(9)   \\
Number of TOAs                                \dotfill &   145       &   141       &   70   \\
Post-fit residual RMS ($\mu$s)                        \dotfill &   136.77     &   49.96    &   34.58  \\
\hline
\multicolumn{4}{c}{Derived Parameters}  \\
\hline\hline
Spin period, $P$ (ms) \dotfill &  51.59136667513(8) &   5.642723444307(3) &   3.254176096627(2) \\
1st Spin period derivative, $\dot{P}$ (s s$^{-1}$) \dotfill &   1.07(2)$\times 10^{-19}$   &   3.431(6)$\times 10^{-20}$  &  $-$1.642(4)$\times 10^{-20}$  \\
\hline
  &        &       &     \\
\hline
\textbf{Pulsar}  &   \textbf{J1801--0857K}   &   \textbf{J1801--0857L}  &   \textbf{J1801--0857M}   \\
\hline\hline
Right ascension, $\alpha$ (J2000)            \dotfill & 18:01:50.7212(9)   &  18:01:49.5847(7)    &    18:01:50.6297(4)  \\
Declination, $\delta$ (J2000)                \dotfill &  --08:57:32.56(4)  &   --08:57:08.06(4)     &    --08:57:31.45(2)   \\
Spin frequency, $f$ (s$^{-1}$)   \dotfill &   104.2699714144(2)  &   165.0902463911(2)  &   186.6773548929(1)  \\
1st Spin frequency derivative, $\dot{f}$ (Hz s$^{-2}$)   \dotfill &   $-$8.893(4)$\times 10^{-15}$  &   9.21(6)$\times 10^{-16}$    &   $-$1.6932(3)$\times 10^{-14}$   \\
Reference epoch (MJD)        \dotfill &   58868.079260   &   58867.093100     &   58871.058500    \\
Dispersion measure, DM (pc cm$^{-3}$)    \dotfill &   182.38(2)   &   185.74(2)  &   183.18(1)    \\
Number of TOAs                           \dotfill &   54           &   60          &   59            \\
Post-fit residual RMS ($\mu$s)                   \dotfill &   108.18       &   93.82       &   55.52         \\
\hline
\multicolumn{4}{c}{Derived Parameters}  \\
\hline\hline
Spin period, $P$ (ms)   \dotfill &   9.59048886688(2)  &   6.057293037354(9)    &   5.356836133518(4)   \\
1st Spin period derivative, $\dot{P}$ (s s$^{-1}$)   \dotfill &   8.180(3)$\times 10^{-19}$  &   $-$3.38(2)$\times 10^{-20}$    &   4.8586(9)$\times 10^{-19}$   \\
\hline
  &        &       &     \\
\hline
\textbf{Pulsar}  &   \textbf{J1801--0857N}    &   \textbf{J1801--0857O}   \\
\hline\hline
Right ascension, $\alpha$ (J2000)            \dotfill & 18:01:50.6318(5)   & 18:01:50.7720(3)    &      \\
Declination, $\delta$ (J2000)                \dotfill &  --08:57:34.50(2)  &   --08:57:33.37(2)     &     \\
Spin frequency, $f$ (s$^{-1}$)                         \dotfill &   200.2170232280(2)          &   233.2571878284(2)               \\
1st Spin frequency derivative, $\dot{f}$ (Hz s$^{-2}$) \dotfill &   2.1306(5)$\times 10^{-14}$ &   $-$7.149(4)$\times 10^{-15}$   \\
Reference epoch (MJD)                                  \dotfill &   58868.079200               &   59941.134200                   \\
Dispersion measure, DM (pc cm$^{-3}$)                  \dotfill &   182.642(9)                 &   182.49(1)                    \\
Number of TOAs                                         \dotfill &   57                         &   54                             \\
Post-fit residual RMS ($\mu$s)                                 \dotfill &   68.11                      &   46.74                          \\
\hline
\multicolumn{4}{c}{Derived Parameters}  \\
\hline\hline
Spin period, $P$ (ms) \dotfill &   4.994580300304(5)  &   4.287113333183(4)  \\
1st Spin period derivative, $\dot{P}$ (s s$^{-1}$)    \dotfill &   $-$5.315(1)$\times 10^{-19}$   &   1.3139(8)$\times 10^{-19}$  \\
\hline
\end{longtable}

\begin{longtable}{l c c c}
\caption{The information of NGC~6517P, Q, R from preliminary timing measurements made to date. 
The timing data spans from MJD 58689, 58867 and 58856 to 59944 for NGC~6517P, Q and R, respectively.
The optical center of NGC~6517 is used as the pulsar's position. The $\dot{f}$ of each pulsar is fixed to 0.
\label{tab:2}} \\
\hline
\endfirsthead %
\multicolumn{4}{c}{} \\ %
\multicolumn{4}{c}{{\tablename\ \thetable{} -- Continued from previous page}} \\
\hline
\endhead %
\hline
\multicolumn{4}{r}{{Continued on next page}} \\
\endfoot %
\hline
\endlastfoot %
\hline
\textbf{Pulsar}  &   \textbf{J1801--0857P} &   \textbf{J1801--0857Q} &   \textbf{J1801--0857R} \\
\hline\hline
Spin frequency, $f$ (s$^{-1}$)       \dotfill &   180.632266(2)    &   137.776526(2)   &   151.8545784(9)    \\
Reference epoch (MJD)  \dotfill &   59944.125910  &   58871.058570   &   59944.125910 \\
Dispersion measure, DM (pc cm$^{-3}$)  \dotfill &   183.05 &   182.45 &   182.50 \\
Number of TOAs           \dotfill &   33   &   24  &   29  \\
Post-fit residual RMS ($\mu$s)    \dotfill &   119.84  &   91.53  &   56.03 \\
\hline
\multicolumn{4}{c}{Derived Parameters}  \\
\hline\hline
Spin period, $P$ (ms)  \dotfill &   5.53610949(6)   &   7.25813048(9)   &   6.58524762(4)   \\
\hline
\end{longtable}

Whilst obtaining or updating the timing solutions for all known pulsars in (NGC~6517A to I) we find solutions
that are consistent with previous studies \citep{2011ApJ...734...89L, 2021RAA....21..143P}, 
albeit with measurements of higher precision 
(e.g., $\alpha$ and $\delta$ were improved in their accuracy by at least one decimal place).
Of the newly discovered pulsars, we can only obtain the phase-connected timing solutions for NGC~6517K, L, M, N, and O.
The phase-connected timing solutions of NGC~6517A to M were obtained by manually removing the JUMPs between different observations.
For the pulsars NGC~6517N and O, the \texttt{Dracula} algorithm returned single solutions with a reduced $\chi^2 < 2$.
Presently, only NGC~6517B is confirmed as a binary pulsar and the other pulsars appear to be isolated.
Updated timing solutions for all previously known pulsars in NGC~6517, in addition to solutions for the newly discovered pulsars described in this work, are presented in Table~\ref{tab:1}.
For the other three pulsars, NGC~6517P, Q, and R, their low flux density (S$_{\rm mean} \sim 1~\mu$Jy) led to intermittent detection through standard search methods. However, once the observations were folded with a basic ephemeris containing the spin period, and adopting the central location of NGC~6517 as the pulsar position, detections could be made in all observations longer than $\sim2\,{\rm h}$. 
Continuous monitoring to increase the number of detections of PSRs~NGC~6517P, Q, and R is needed to remove JUMPs (see Table~2 and Figure~1). 
Since the submission of this letter,
three weaker millisecond pulsars, NGC~6517S to U, were detected in this cluster.
Their spin periods are 3.77 ms, 3.68 ms, and 6.02 ms, respectively, and more details will be discussed in another works (Dai et al. in preparation).

Our fits for DM, including the newly discovered pulsars, show that the difference between the maximum and minimum DM of pulsars in NGC~6517 is 
$\sim$11.2 cm$^{-3}$\,pc, 
making it the  GC with the second largest spread in pulsar DMs.
Only GLIMPSE-C01, with pulsars that show a DM range of $\sim$62 cm$^{-3}$\,pc, is larger.
The DM difference may be caused by the intrinsic nature of the interstellar medium (ISM) within  GCs, or the turbulent nature of the Galactic ISM on small angular scales (see e.g., \citealt{2005ApJ...621..959F}). NGC~6517, located close to the Galactic plane, is therefore a useful yardstick for understanding these effects. 
Three X-ray sources were detected in 0.3-8\,keV with luminosities ranging from $2.7 \times 10^{31}$ erg s$^{-1}$ to $2.4 \times 10^{32}$ erg s$^{-1}$ \citep{2022MNRAS.511.5964Z}.
Because nine of the 17 pulsars lie within the core-radius,
their possible association with X-ray counterparts will be discussed in another paper (Yin et al. in preparation).  

\section{An update on the estimated number of GC pulsars}
\label{psr_numbsers}

After making the first detections of pulsars in NGC~6517, \citet{2011ApJ...734...89L} made complete population predictions for this GC using three observed pulsar luminosity functions, and with limits defined by the peak luminosity of the pulsars detected and the weakest known GC pulsar at that time. They found that 12 to 17 pulsars would exist in this GC. We note the upper limit from \citet{2011ApJ...734...89L} now matches the known pulsar population in NGC~6517, however, as discussed below, the new pulsar discoveries described in this work change population predictions. Since then other methods have been employed that not only predict the number of pulsars in GCs such as NGC~6517, but try to establish possible correlation between the intrinsic cluster properties and their pulsar count \citep{2013MNRAS.436.3720T}. The empirical Bayesian method described by \citet{2013MNRAS.436.3720T} was designed to model pulsar populations that are heavily biased by selection effects resulting in populations with either low to zero pulsar detections. For NGC~6517 this method results in a prediction of $\sim15$ pulsars. 

Using the publicly available software \footnote{\url{http://astro.phys.wvu.edu/gcpsrs/empbayes}} described in \citet{2013MNRAS.436.3720T} it was found that by incorporating recent pulsar discoveries from FAST, the predicted number of pulsars using this method in GCs M53 and M71 differed from earlier estimations by 3 to 14 and 1 to 4 pulsars respectively. 
Because all GCs in the FAST sky have now been searched for undiscovered pulsars, resulting in a total of 50 new discoveries\footnote{\url{https://fast.bao.ac.cn/cms/article/65}}, in this work we used the updated detected pulsar count, including FAST detections, to re-evaluate the optimal models found using the method described by \citet{2013MNRAS.436.3720T}.
As this method can be used to investigate the largely unknown relationship between GC properties and their pulsar population, incorporating the increased pulsar sample is expected to elucidate this further.

We applied largely the same methods as outlined in \citet{2013MNRAS.436.3720T}, and refer the reader to this work for detailed descriptions of the statistical methods used. In the following, only the necessary definitions have been explained. The GC parameters to be investigated included the optical $V$-band luminosity ($L_{\rm V}$); escape velocity ($V_{\rm esc}$); metallicities ($M_{\rm L}$) from \citealt{1996AJ....112.1487H} (2010 edition); stellar encounter rates ($\Gamma$) from \citet{2013ApJ...766..136B};
and minimum detectable flux densities ($S_{\rm min}$) from \citet{2011ApJ...742...51B}. In \citet{2013MNRAS.436.3720T} a total of 94 GCs (with 144 known pulsars in 28 of the GCs) with all five of these parameters measured were used. An a-priori candidate set of 240 models, $\hat{N}_{\rm psr, i}$, were modelled with different covariate effects, $X_{ij}$, which are combinations of the physical properties of the GCs mentioned above and their regression coefficients, $\beta_{i}$. Here $X_{ij}$ is the $j^{\rm th}$ covariate in $i^{\rm th}$ GC ($j = 1, 2, . . . , r$) and the regression coefficient $\beta_{j}$ characterises the effect of the $j^{\rm th}$ covariate. These were then evaluated by the Akaike's Information Criterion \citep[AIC,][]{burnham2004model} to select optimal models; where models with smaller AIC values (by at least two AIC units, \citealt{burnham2004model}) are considered to provide a better fit to the sample data. 
In \citet{2013MNRAS.436.3720T} the estimated detection probability, $\hat{p}_{i}$, for the $i^{\rm th}$ GC was considered in three prior distributions on GC pulsar population ($\hat{N}_{\rm psr, i}$):
Poisson (P), negative binomial (NB), and zero-inflated Poisson (ZIP). Through their analysis \citet{2013MNRAS.436.3720T} found that the stellar encounter rate is the most critical parameter in estimating the number of pulsars in a GC.

In our updated analysis we divided the GC samples into two groups. Firstly we fitted models using all 94 GCs, but with updated pulsar counts. Our second sample consisted of only the GCs visible from FAST. The reasoning for this is described below.
For the first sample of all 94 GCs, 
the results from the AIC ranking showed that models which included escape velocity were favored over models in which this was not a factor.
The results from the most simplified models (i.e., those with the fewest parameters) within 2 AIC units are shown in the upper section of Table~\ref{tab:3}.
The second sample of FAST only GCs was chosen for the following reasons.  To date, almost all GCs within the FAST sky have been deeply searched for pulsars (limiting flux of $\sim1\,\mu{\rm Jy}$, \citealt{2021ApJ...915L..28P}) by FAST. The searches for GC pulsars outside the FAST sky have been carried out by different telescopes with different sensitivity and observing systems. Therefore we assume that a sample with  uniform sensitivity coverage could reduce the influence of selection bias and more accurately reveal the underlying  distribution of the pulsar population in GCs.

\begin{figure}
\centering
\includegraphics[width=12cm, angle=0]{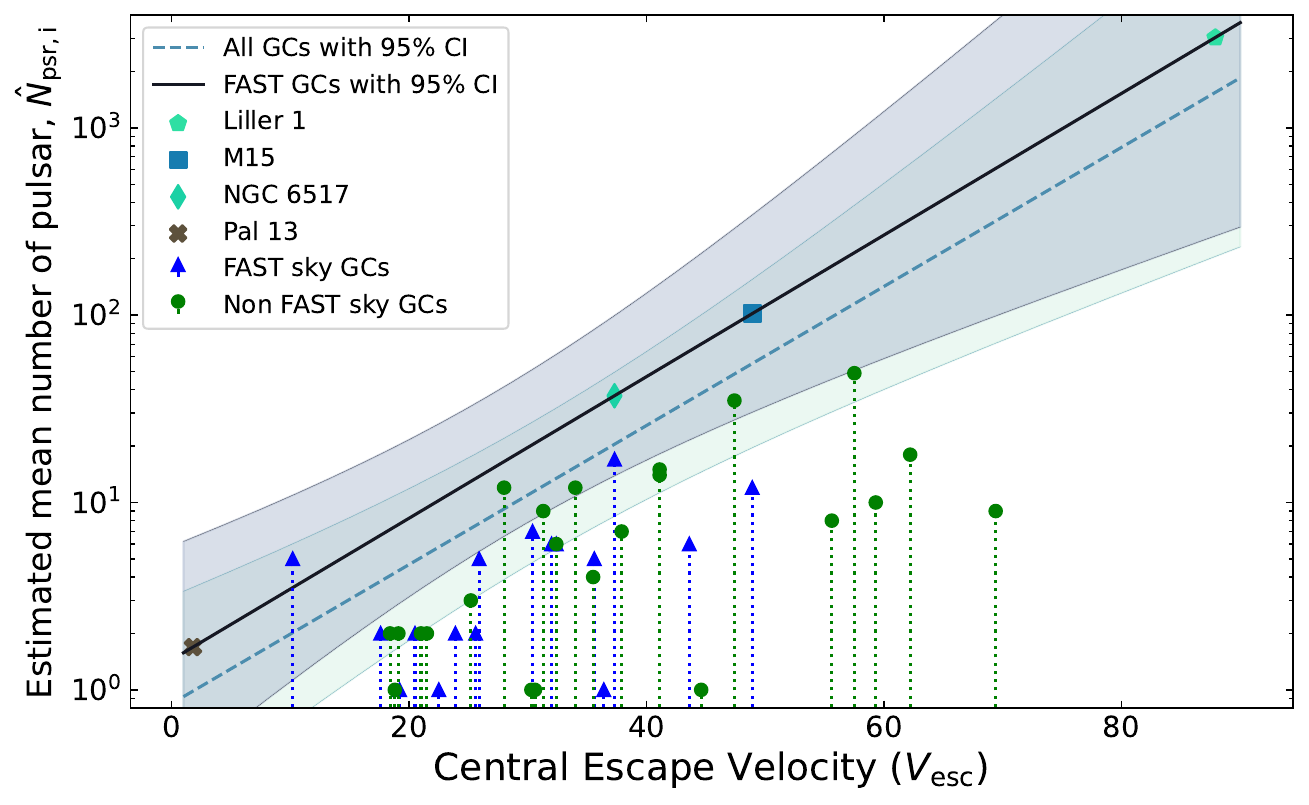}
\caption{The correlation between the number of pulsars and the central escape velocity of Galactic GCs.
The lines with 95$\%$ credible interval (CI) in grey colour shows the relation between the escape velocity and the number of estimated pulsars, 
while the stem plot shows the escape velocity of 39 GCs with known pulsars and the number of known pulsars in them.
The blue dotted line and the black solid line present the model derived from all GCs and only GCs within FAST sky, respectively.
GLIMPSE-C01 (with 6 known pulsars) was not included in the figure as its escape velocity was not available.
In the sample, Pal~13 and M15 are the minimum and maximum escape velocities in the FAST sky, respectively.
Liller 1 has the largest escape velocity in the Galactic GCs.
}
\label{Fig2}
\end{figure}

There are 45 GCs in the FAST sky and 
amongst these, 31 were used because they have all five of the necessary GC property parameters. 
Results from the three models most highly ranked by their AIC values are presented in the lower section of 
Table~\ref{tab:3}. 
For our GC sample covering the FAST sky, models which included the GC escape velocity $V_{\rm esc}$ are also favored over other models in which this is not a factor.
The AIC values of the three models in the lower section of Table~\ref{tab:3} are almost equal. 
Here, in order to further briefly compare the difference between the results of the two samples in the estimated counts of pulsars for all GCs in the Milky Way, we used the optimal model with only one intercept term for further analysis since the estimated detection probabilities of pulsars in some GCs
were not available.
We therefore adopted the second model (Model no. 2 in the lower section of Table~\ref{tab:3}), with an AIC value of $103.88$, and in which $\beta_0=0.369$ and $\beta_1=0.087$, for further study. 
 The association from the selected model is significant with a $P$-value\footnote{The probability, $P$-value, is a measure of evidence against the null hypothesis in favor of the alternative hypothesis, i.e., a smaller $P$-value indicates more significant the correlation between the parameter $V_{\rm esc}$ and $\hat{N}_{\rm psr}$ (see \citealt{2013MNRAS.436.3720T} and reference therein for details).} of $5.94\times10^{-6} < 0.05$.
This provides further support for the potential correlation between the population of pulsars in GCs and their escape velocities. 
For the sake of brevity (i.e., only one intercept term), the third model from the sample of all GCs was used for further comparisons with the selected model from the FAST sky.
The association of the selected model from all GCs is also significant with a $P$-value of $1.78\times10^{-7}$ ($\beta_0=-0.169$, $\beta_1=0.086$, and AIC=314.60).
Assuming that the correlation from the GCs within the FAST is applicable to Galactic GCs outside of the FAST sky, 
the pulsar numbers of these GCs were estimated accordingly (see Figure~\ref{Fig2} and Table~\ref{table:longtable_example_1}).
As seen in Figure~\ref{Fig2}, the estimated counts between the FAST sky model and all sky model are slightly different, which may be due to the higher pulsar detection rate in GCs from the FAST (e.g., \citealt{2023RAA....23e5012Y}).

\begin{table}
\caption{The top models within two AIC units to the smallest AIC. $\Delta$AIC is the change in AIC relative to the model with smallest AIC value and Type refers to the model prior used for abundance, in this case Negative Binomial (NB), see \citet{2013MNRAS.436.3720T} for details.
The models in the upper section are from all GCs, and the results in the lower section are from GCs within the FAST sky.
Note the AIC value can only be used to rank models within the same sample, not between different samples.
}
\label{tab:3}
\begin{center}
\begin{tabular}{clllll}
\hline
 &  The sample of all GCs &   &   &  \\  
\hline
\multicolumn{1}{l}{Model no.} & {Model Structure} & AIC & $\Delta$AIC      & Type        \\
\hline
1. & $\ln\hat{N}_{\rm psr, i} $ = $\beta_0 + \beta_{1} V_{\rm esc} + \lg(\widehat{p}_i)   $   & 313.65 & 0.37  &   NB    \\
2. & $\ln\hat{N}_{\rm psr, i} $ = $\beta_0 + \beta_{1} V_{\rm esc} + \widehat{p}_i $          & 314.16 & 0.88  &   NB   \\
3. & $\ln\hat{N}_{\rm psr, i} $ = $\beta_0 + \beta_{1} V_{\rm esc}$                           & 314.60 & 1.32  &   NB   \\
\hline
\hline
& \multicolumn{1}{c}{The sample of only GCs within FAST sky.}   \\
\hline
1. & $\ln\hat{N}_{\rm psr, i} $ = $\beta_0 + \beta_{1} V_{\rm esc} + \widehat{p}_i $       & 103.84 & 0.000  &   NB   \\
2. & $\ln\hat{N}_{\rm psr, i} $ = $\beta_0 + \beta_{1} V_{\rm esc}$                        & 103.88 & 0.035  &   NB   \\
3. & $\ln\hat{N}_{\rm psr, i} $ = $\beta_0 + \beta_{1} V_{\rm esc} + \lg(\widehat{p}_i)  $  & 103.91 & 0.063  &   NB    \\
\hline
\end{tabular}
\end{center}
\end{table}


The fact that GCs with larger total mass also have, on average, larger escape velocities is a natural correlation.
The $V_{\rm esc}$ was derived from the best-fitting N-body models of each GC in Baumgardt's list, which can be estimated by $V_{\rm esc}=\sqrt{2GM/r_{\rm h,l}}$ or $V_{\rm esc}=\sqrt{1.5GM/r_{\rm h,m}}$, where M, $r_{\rm h,l}$, and $r_{\rm h,m}$ are the GC mass, projected half-light radius, and half-mass radius (3D), respectively.
Because neutron stars receive a high natal kick velocity (e.g., \citealt{2005MNRAS.360..974H}) it was generally predicted that they would not be retained within GCs, 
and therefore 
those within GCs have been proposed to form through a low-velocity channel (e.g., electron-capture supernovae, \citealt{2008MNRAS.386..553I, 2013IAUS..291..243F}).
Assuming that the evolutionary nature of Galactic GCs 
are alike, simply the total mass of the GCs may influence the number of 
neutron stars originally retained, as pointed out by \citet{2008AIPC..983..415R}.
Therefore, more initial neutron stars can be formed due to a larger cluster mass, while a sufficiently high escape velocity of the cluster considerably ensures retention of the neutron stars in this stage (e.g., \citealt{2008MNRAS.386..553I}). 
After retention, the initial neutron stars may be recycled to millisecond spin periods by accretion.
The escape velocity of cluster is related to the total mass and size of the cluster (i.e. related to density).
Higher central density of GC leads to an increase in the formation of more low-mass X-ray binaries and their offspring of millisecond pulsars through close stellar encounters (e.g., \citealt{1982Natur.300..728A, 2003ApJ...591L.131P}).
The dependence here between the number of pulsars and the escape velocity 
is consistent with previous N-body simulation works exploring the formation of millisecond pulsars in GCs (e.g., \citealt{2008MNRAS.386..553I, 2019ApJ...877..122Y}).

The range of $V_{\rm esc}$ of the 31 GCs in the FAST sky is $1.8\,{\rm km\,s}^{-1}$ (Pal 13) to $48.9\,{\rm km\,s}^{-1}$ (M15). In Table~\ref{table:longtable_example} the results of our pulsar number predictions, and those from \citet{2013MNRAS.436.3720T}  are given in order of descending $V_{\rm esc}$. Thus, in the FAST sky M15 (13 pulsars, M15A to M) was expected to host 102 pulsars, 
being the GC with the largest number of predicted pulsars.
Predicted to host 64 pulsars and
with at least 6 pulsars discovered by FAST (e.g., \citealt{2021ApJ...915L..28P}), 
M2 is suggested to be the GC with the second largest number of pulsars.
With only one pulsar, an eclipsing redback namely M92A, 
discovered in 2017 by FAST \citep{2020ApJ...892L...6P},
the modeled pulsar number for M92 is 34.
Thus, we suggest that more pulsars should be discovered in M92.
On the other hand, the possible reasons of missing pulsars can be
either the relatively low fluxes and or compact orbits.
The only exception is M71.
There are 5 pulsars in it, while the modeled pulsar number is 4.

With the $V_{\rm esc}$ of 87.9 km s$^{-1}$, Liller 1 is predicted to host thousands of pulsars.
The prediction is supported by the
previous estimation that 
it could host 410$_{-210}^{+480}$ pulsars (\citealt{2010A&A...524A..75A, 2011ApJ...729...90T}) due to its highest $\gamma$-ray luminosity among all GCs.
The distance of this GC from the Sun is 8.2\,kpc, 
close enough for finding new pulsars.
Located in the bulge of the Milky Way, the heavy dispersive smearing and scatter broadening 
caused by possible high DM values (e.g., $\sim$ 697 cm$^{-3}$ pc from NE2001, \citealt{2002astro.ph..7156C}; $\sim$ 1104 cm$^{-3}$ pc from YMW16, \citealt{2017ApJ...835...29Y})
can be the reason for missing pulsars.
The pulsars in this GC 
might require higher frequency bands at large radio telescopes.

M54, NGC~6388, NGC~2808, and M75 are the other four GCs with similar or higher $V_{\rm esc}$ than M15, but with few pulsars detected.
M54 and M75 are located more than 20\,kpc away, further than M53 which is the furthest GC with known pulsars. 
The distance to either NGC~6388 or NGC~2808 is around 10\,kpc, similar to Liller1. 
So, we suggest that the priority of pulsar surveys towards NGC~6388 and NGC~2808 should be higher than those to M54 and M75.

We caution against directly applying the above estimated numbers, as selection bias may still exist and a further analysis is required to refine this simulation.
The flux density limit used here \citep{2011ApJ...742...51B}, as discussed in \citet{2013MNRAS.436.3720T}, is strictly applicable only to long-period pulsars, and thus we focus here on the potential correlation between the pulsar population of GCs and their physical parameters, without analyzing the pulsar detection probability.
The updating of some of those numbers may refine the analysis.
On the premise that almost all GCs within the FAST sky have been deeply searched for pulsars at the same sensitivity limit, we assume that the population of pulsars within these GCs is more akin to the native population, but a selection bias may still exist by using only these GCs as a sample to estimate the whole Galactic population distribution.
For example, most GCs are located around the Galactic center or in the Southern sky, where some of them host the largest number of pulsars, while the GCs within the FAST sky are mostly out of the Galactic plane.
Also, most GCs in the FAST sky have smaller amounts of DM and scattering, enabling the initial pulsars within them to be more easily detected.
Similarly, the slight difference between the estimated counts derived from the FAST sky and all sky may also arise from the different ranges of cluster physics parameters, such as escape velocities.

In the FAST sky, around 50 GC pulsars were discovered by either FAST (e.g., \citealt{2021ApJ...915L..28P}) or others (e.g., Glimpse~C01 A, \citealt{2023arXiv231211694M}).
With known pulsars, the computational time cost for dedisperion can be largely saved.
Thus, those GCs with known pulsars should be prioritized for future observing campaigns.
After searching for almost all the GCs in FAST sky, 
20 pulsars were discovered in the GCs with no previously known pulsars, 
including M92, M14, M2, NGC~6712, M10, and M12.
The probability of pulsar discoveries in GCs with or without known pulsars may not differ significantly. 
On the other hand, short observations should be carried out before longer ones to reduce unnecessary processing time and to improve chances of finding compact binaries. 
Based on predictions made here, there is little doubt that the ongoing and future GC pulsar surveys with GMRT (e.g., \citealt{2022A&A...664A..54G}), Parkes (e.g., \citealt{2022ApJ...934L..21Z}), MeerKAT (e.g., \citealt{2021MNRAS.504.1407R}), and SKA \citep{2015aska.confE..47H} in the near future will bring more discoveries.

\section{Conclusions}
\label{Conclusions}

In this study of FAST observations of NGC 6517, we have reached the following conclusions:
\begin{enumerate}
\item{Eight new isolated millisecond pulsars, namely PSR~J1801$-$0857K to R, or, NGC~6517K to R, were discovered in NGC~6517.
Up to the end of 2023, there are 17 pulsars detected in NGC~6517.}

\item{NGC~6517J is proven to be a harmonic of NGC~6517B. 
The phase-connected timing solutions of NGC~6517K to O were obtained. The timing solutions of previously known pulsars were updated and are consistent with previous works.}

\item{In the framework of the empirical Bayesian model used in this paper, the recently observed sample of GC pulsars favored a correlation between the number of potential pulsars and the central escape velocities unlike the original prediction of \citet{2013MNRAS.436.3720T} using the same method which predicted the stellar encounter rate as the main factor.}

\item{In the FAST sky, M2 and M92 may have more pulsars.
Outside of the FAST sky, Liller 1, NGC~6388, NGC~2808, M54, and M75 have the highest possibilities for finding new pulsars.} 
\end{enumerate}

\begin{acknowledgments}

This work is supported by National SKA Program of China (No. 2020SKA0120100),  the Basic Science Center Project of the National Nature Science Foundation of China (NSFC) under Grant Nos. 11703047, 11773041, U1931128, U2031119, 12003047 and 12173053.
Lei Qian is supported by the Youth Innovation Promotion Association of CAS (id.~2018075, Y2022027), and the CAS ``Light of West China'' Program. RPE is supported by the Chinese Academy of Sciences President's International Fellowship Initiative, Grant No. 2021FSM0004.
M.L. is supported by Guizhou Provincial Basic Research Program (Natural Science) (ZK[2023] 039), Key Technology R\&D Program([2023] 352) and National Natural Science Foundation of China under Grand No. 12363010. 
This work made use of the data from FAST (Five-hundred-meter Aperture Spherical radio Telescope).
FAST is a Chinese national mega-science facility, operated by National Astronomical Observatories, Chinese Academy of Sciences.
This work is also supported by the International Centre of Supernovae, Yunnan Key Laboratory (No. 202302AN360001 and 202302AN36000104).
Finally, we thank the anonymous referees for their helpful 
suggestions to bring clarity to the text.

\end{acknowledgments}

\bibliography{NGC6517_Pulsars}{}

\begin{thebibliography}{}
\expandafter\ifx\csname natexlab\endcsname\relax\def\natexlab#1{#1}\fi
\providecommand{\url}[1]{\href{#1}{#1}}
\providecommand{\dodoi}[1]{doi:~\href{http://doi.org/#1}{\nolinkurl{#1}}}
\providecommand{\doeprint}[1]{\href{http://ascl.net/#1}{\nolinkurl{http://ascl.net/#1}}}
\providecommand{\doarXiv}[1]{\href{https://arxiv.org/abs/#1}{\nolinkurl{https://arxiv.org/abs/#1}}}

\bibitem[{{Abdo} {et~al.}(2009){Abdo}, {Ackermann}, {Ajello}, {Atwood},
  {Axelsson}, {Baldini}, {Ballet}, {Barbiellini}, {Baring}, {Bastieri},
  {Baughman}, {Bechtol}, {Bellazzini}, {Berenji}, {Bignami}, {Blandford},
  {Bloom}, {Bonamente}, {Borgland}, {Bregeon}, {Brez}, {Brigida}, {Bruel},
  {Burnett}, {Caliandro}, {Cameron}, {Camilo}, {Caraveo}, {Carlson},
  {Casandjian}, {Cecchi}, {{\c{C}}elik}, {Charles}, {Chekhtman}, {Cheung},
  {Chiang}, {Ciprini}, {Claus}, {Cognard}, {Cohen-Tanugi}, {Cominsky},
  {Conrad}, {Corbet}, {Cutini}, {Dermer}, {Desvignes}, {de Angelis}, {de Luca},
  {de Palma}, {Digel}, {Dormody}, {do Couto e Silva}, {Drell}, {Dubois},
  {Dumora}, {Edmonds}, {Farnier}, {Favuzzi}, {Fegan}, {Focke}, {Frailis},
  {Freire}, {Fukazawa}, {Funk}, {Fusco}, {Gargano}, {Gasparrini}, {Gehrels},
  {Germani}, {Giebels}, {Giglietto}, {Giordano}, {Glanzman}, {Godfrey},
  {Grenier}, {Grondin}, {Grove}, {Guillemot}, {Guiriec}, {Hanabata}, {Harding},
  {Hayashida}, {Hays}, {Hobbs}, {Hughes}, {J{\'o}hannesson}, {Johnson},
  {Johnson}, {Johnson}, {Johnson}, {Johnston}, {Kamae}, {Katagiri}, {Kataoka},
  {Kawai}, {Kerr}, {Kn{\"o}dlseder}, {Kocian}, {Kramer}, {Kuss}, {Lande},
  {Latronico}, {Lemoine-Goumard}, {Longo}, {Loparco}, {Lott}, {Lovellette},
  {Lubrano}, {Madejski}, {Makeev}, {Manchester}, {Marelli}, {Mazziotta},
  {McConville}, {McEnery}, {McLaughlin}, {Meurer}, {Michelson}, {Mitthumsiri},
  {Mizuno}, {Moiseev}, {Monte}, {Monzani}, {Morselli}, {Moskalenko}, {Murgia},
  {Nolan}, {Norris}, {Nuss}, {Ohsugi}, {Omodei}, {Orlando}, {Ormes}, {Paneque},
  {Panetta}, {Parent}, {Pelassa}, {Pepe}, {Pesce-Rollins}, {Piron}, {Porter},
  {Rain{\`o}}, {Rando}, {Ransom}, {Ray}, {Razzano}, {Rea}, {Reimer}, {Reimer},
  {Reposeur}, {Ritz}, {Rochester}, {Rodriguez}, {Romani}, {Roth}, {Ryde},
  {Sadrozinski}, {Sanchez}, {Sander}, {Saz Parkinson}, {Scargle}, {Schalk},
  {Sgr{\`o}}, {Siskind}, {Smith}, {Smith}, {Spandre}, {Spinelli}, {Stappers},
  {Starck}, {Striani}, {Strickman}, {Suson}, {Tajima}, {Takahashi}, {Tanaka},
  {Thayer}, {Thayer}, {Theureau}, {Thompson}, {Thorsett}, {Tibaldo}, {Torres},
  {Tosti}, {Tramacere}, {Uchiyama}, {Usher}, {Van Etten}, {Vasileiou},
  {Venter}, {Vilchez}, {Vitale}, {Waite}, {Wallace}, {Wang}, {Watters}, {Webb},
  {Weltevrede}, {Winer}, {Wood}, {Ylinen}, \& {Ziegler}}]{2009Sci...325..848A}
{Abdo}, A.~A., {Ackermann}, M., {Ajello}, M., {et~al.} 2009, Science, 325, 848,
  \dodoi{10.1126/science.1176113}

\bibitem[{{Abdo} {et~al.}(2010){Abdo}, {Ackermann}, {Ajello}, {Baldini},
  {Ballet}, {Barbiellini}, {Bastieri}, {Bellazzini}, {Blandford}, {Bloom},
  {Bonamente}, {Borgland}, {Bouvier}, {Brandt}, {Bregeon}, {Brigida}, {Bruel},
  {Buehler}, {Buson}, {Caliandro}, {Cameron}, {Caraveo}, {Carrigan},
  {Casandjian}, {Charles}, {Chaty}, {Chekhtman}, {Cheung}, {Chiang}, {Ciprini},
  {Claus}, {Cohen-Tanugi}, {Conrad}, {Decesar}, {Dermer}, {de Palma}, {Digel},
  {Silva}, {Drell}, {Dubois}, {Dumora}, {Favuzzi}, {Fortin}, {Frailis},
  {Fukazawa}, {Fusco}, {Gargano}, {Gasparrini}, {Gehrels}, {Germani},
  {Giglietto}, {Giordano}, {Glanzman}, {Godfrey}, {Grenier}, {Grondin},
  {Grove}, {Guillemot}, {Guiriec}, {Hadasch}, {Harding}, {Hays}, {Jean},
  {J{\'o}hannesson}, {Johnson}, {Johnson}, {Kamae}, {Katagiri}, {Kataoka},
  {Kerr}, {Kn{\"o}dlseder}, {Kuss}, {Lande}, {Latronico}, {Lee},
  {Lemoine-Goumard}, {Llena Garde}, {Longo}, {Loparco}, {Lovellette},
  {Lubrano}, {Makeev}, {Mazziotta}, {Michelson}, {Mitthumsiri}, {Mizuno},
  {Monte}, {Monzani}, {Morselli}, {Moskalenko}, {Murgia}, {Naumann-Godo},
  {Nolan}, {Norris}, {Nuss}, {Ohsugi}, {Omodei}, {Orlando}, {Ormes},
  {Pancrazi}, {Parent}, {Pepe}, {Pesce-Rollins}, {Piron}, {Porter},
  {Rain{\`o}}, {Rando}, {Reimer}, {Reimer}, {Reposeur}, {Ripken}, {Romani},
  {Roth}, {Sadrozinski}, {Saz Parkinson}, {Sgr{\`o}}, {Siskind}, {Smith},
  {Spinelli}, {Strickman}, {Suson}, {Takahashi}, {Takahashi}, {Tanaka},
  {Thayer}, {Thayer}, {Tibaldo}, {Torres}, {Tosti}, {Tramacere}, {Uchiyama},
  {Usher}, {Vasileiou}, {Venter}, {Vilchez}, {Vitale}, {Waite}, {Wang}, {Webb},
  {Winer}, {Yang}, {Ylinen}, {Ziegler}, \& {Fermi LAT
  Collaboration}}]{2010A&A...524A..75A}
---. 2010, \aap, 524, A75, \dodoi{10.1051/0004-6361/201014458}

\bibitem[{{Alpar} {et~al.}(1982){Alpar}, {Cheng}, {Ruderman}, \&
  {Shaham}}]{1982Natur.300..728A}
{Alpar}, M.~A., {Cheng}, A.~F., {Ruderman}, M.~A., \& {Shaham}, J. 1982, \nat,
  300, 728, \dodoi{10.1038/300728a0}

\bibitem[{{Anderson} {et~al.}(1990){Anderson}, {Gorham}, {Kulkarni}, {Prince},
  \& {Wolszczan}}]{1990Natur.346...42A}
{Anderson}, S.~B., {Gorham}, P.~W., {Kulkarni}, S.~R., {Prince}, T.~A., \&
  {Wolszczan}, A. 1990, \nat, 346, 42, \dodoi{10.1038/346042a0}

\bibitem[{{Bahramian} {et~al.}(2013){Bahramian}, {Heinke}, {Sivakoff}, \&
  {Gladstone}}]{2013ApJ...766..136B}
{Bahramian}, A., {Heinke}, C.~O., {Sivakoff}, G.~R., \& {Gladstone}, J.~C.
  2013, \apj, 766, 136, \dodoi{10.1088/0004-637X/766/2/136}

\bibitem[{{Barr} {et~al.}(2024){Barr}, {Dutta}, {Freire}, {Cadelano}, {Gautam},
  {Kramer}, {Pallanca}, {Ransom}, {Ridolfi}, {Stappers}, {Tauris}, {Venkatraman
  Krishnan}, {Wex}, {Bailes}, {Behrend}, {Buchner}, {Burgay}, {Chen},
  {Champion}, {Chen}, {Corongiu}, {Geyer}, {Men}, {Padmanabh}, \&
  {Possenti}}]{2024Sci...383..275B}
{Barr}, E.~D., {Dutta}, A., {Freire}, P. C.~C., {et~al.} 2024, Science, 383,
  275, \dodoi{10.1126/science.adg3005}

\bibitem[{{Blandford} \& {Teukolsky}(1976)}]{1976ApJ...205..580B}
{Blandford}, R., \& {Teukolsky}, S.~A. 1976, \apj, 205, 580,
  \dodoi{10.1086/154315}

\bibitem[{{Boyles} {et~al.}(2011){Boyles}, {Lorimer}, {Turk}, {Mnatsakanov},
  {Lynch}, {Ransom}, {Freire}, \& {Belczynski}}]{2011ApJ...742...51B}
{Boyles}, J., {Lorimer}, D.~R., {Turk}, P.~J., {et~al.} 2011, \apj, 742, 51,
  \dodoi{10.1088/0004-637X/742/1/51}

\bibitem[{Burnham \& Anderson(2004)}]{burnham2004model}
Burnham, K.~P., \& Anderson, D.~R. 2004, A practical information-theoretic
  approach, 2

\bibitem[{{Camilo} \& {Rasio}(2005)}]{2005ASPC..328..147C}
{Camilo}, F., \& {Rasio}, F.~A. 2005, in Astronomical Society of the Pacific
  Conference Series, Vol. 328, Binary Radio Pulsars, ed. F.~A. {Rasio} \& I.~H.
  {Stairs}, 147, \dodoi{10.48550/arXiv.astro-ph/0501226}

\bibitem[{{Cordes} \& {Lazio}(2002)}]{2002astro.ph..7156C}
{Cordes}, J.~M., \& {Lazio}, T.~J.~W. 2002, arXiv e-prints, astro,
  \dodoi{10.48550/arXiv.astro-ph/0207156}

\bibitem[{{Davies} {et~al.}(1992){Davies}, {Benz}, \&
  {Hills}}]{1992ApJ...401..246D}
{Davies}, M.~B., {Benz}, W., \& {Hills}, J.~G. 1992, \apj, 401, 246,
  \dodoi{10.1086/172056}

\bibitem[{{Freire}(2013)}]{2013IAUS..291..243F}
{Freire}, P. C.~C. 2013, in Neutron Stars and Pulsars: Challenges and
  Opportunities after 80 years, ed. J.~{van Leeuwen}, Vol. 291, 243--250,
  \dodoi{10.1017/S1743921312023770}

\bibitem[{{Freire} {et~al.}(2005){Freire}, {Hessels}, {Nice}, {Ransom},
  {Lorimer}, \& {Stairs}}]{2005ApJ...621..959F}
{Freire}, P. C.~C., {Hessels}, J. W.~T., {Nice}, D.~J., {et~al.} 2005, \apj,
  621, 959, \dodoi{10.1086/427748}

\bibitem[{{Freire} \& {Ridolfi}(2018)}]{2018MNRAS.476.4794F}
{Freire}, P. C.~C., \& {Ridolfi}, A. 2018, \mnras, 476, 4794,
  \dodoi{10.1093/mnras/sty524}

\bibitem[{{Gautam} {et~al.}(2022){Gautam}, {Ridolfi}, {Freire}, {Wharton},
  {Gupta}, {Ransom}, {Oswald}, {Kramer}, \& {DeCesar}}]{2022A&A...664A..54G}
{Gautam}, T., {Ridolfi}, A., {Freire}, P.~C.~C., {et~al.} 2022, \aap, 664, A54,
  \dodoi{10.1051/0004-6361/202243062}

\bibitem[{{Harris}(1996)}]{1996AJ....112.1487H}
{Harris}, W.~E. 1996, \aj, 112, 1487, \dodoi{10.1086/118116}

\bibitem[{{Hessels} {et~al.}(2015){Hessels}, {Possenti}, {Bailes}, {Bassa},
  {Freire}, {Lorimer}, {Lynch}, {Ransom}, \& {Stairs}}]{2015aska.confE..47H}
{Hessels}, J., {Possenti}, A., {Bailes}, M., {et~al.} 2015, in Advancing
  Astrophysics with the Square Kilometre Array (AASKA14), 47,
  \dodoi{10.22323/1.215.0047}

\bibitem[{{Hessels} {et~al.}(2007){Hessels}, {Ransom}, {Stairs}, {Kaspi}, \&
  {Freire}}]{2007ApJ...670..363H}
{Hessels}, J.~W.~T., {Ransom}, S.~M., {Stairs}, I.~H., {Kaspi}, V.~M., \&
  {Freire}, P.~C.~C. 2007, \apj, 670, 363, \dodoi{10.1086/521780}

\bibitem[{{Hobbs} {et~al.}(2005){Hobbs}, {Lorimer}, {Lyne}, \&
  {Kramer}}]{2005MNRAS.360..974H}
{Hobbs}, G., {Lorimer}, D.~R., {Lyne}, A.~G., \& {Kramer}, M. 2005, \mnras,
  360, 974, \dodoi{10.1111/j.1365-2966.2005.09087.x}

\bibitem[{{Hotan} {et~al.}(2004){Hotan}, {van Straten}, \&
  {Manchester}}]{2004PASA...21..302H}
{Hotan}, A.~W., {van Straten}, W., \& {Manchester}, R.~N. 2004, \pasa, 21, 302,
  \dodoi{10.1071/AS04022}

\bibitem[{{Ivanova} {et~al.}(2008){Ivanova}, {Heinke}, {Rasio}, {Belczynski},
  \& {Fregeau}}]{2008MNRAS.386..553I}
{Ivanova}, N., {Heinke}, C.~O., {Rasio}, F.~A., {Belczynski}, K., \& {Fregeau},
  J.~M. 2008, \mnras, 386, 553, \dodoi{10.1111/j.1365-2966.2008.13064.x}

\bibitem[{{Jonas} \& {MeerKAT Team}(2016)}]{2016mks..confE...1J}
{Jonas}, J., \& {MeerKAT Team}. 2016, in MeerKAT Science: On the Pathway to the
  SKA, 1, \dodoi{10.22323/1.277.0001}

\bibitem[{{King} {et~al.}(2001){King}, {Pringle}, \&
  {Wickramasinghe}}]{2001MNRAS.320L..45K}
{King}, A.~R., {Pringle}, J.~E., \& {Wickramasinghe}, D.~T. 2001, \mnras, 320,
  L45, \dodoi{10.1046/j.1365-8711.2001.04184.x}

\bibitem[{{Kremer} {et~al.}(2023){Kremer}, {Fuller}, {Piro}, \&
  {Ransom}}]{2023MNRAS.525L..22K}
{Kremer}, K., {Fuller}, J., {Piro}, A.~L., \& {Ransom}, S.~M. 2023, \mnras,
  525, L22, \dodoi{10.1093/mnrasl/slad088}

\bibitem[{{Kremer} {et~al.}(2022){Kremer}, {Ye}, {K{\i}ro{\u{g}}lu},
  {Lombardi}, {Ransom}, \& {Rasio}}]{2022ApJ...934L...1K}
{Kremer}, K., {Ye}, C.~S., {K{\i}ro{\u{g}}lu}, F., {et~al.} 2022, \apjl, 934,
  L1, \dodoi{10.3847/2041-8213/ac7ec4}

\bibitem[{{Lynch} {et~al.}(2011){Lynch}, {Ransom}, {Freire}, \&
  {Stairs}}]{2011ApJ...734...89L}
{Lynch}, R.~S., {Ransom}, S.~M., {Freire}, P. C.~C., \& {Stairs}, I.~H. 2011,
  \apj, 734, 89, \dodoi{10.1088/0004-637X/734/2/89}

\bibitem[{{Lyne} {et~al.}(1987){Lyne}, {Brinklow}, {Middleditch}, {Kulkarni},
  \& {Backer}}]{1987Natur.328..399L}
{Lyne}, A.~G., {Brinklow}, A., {Middleditch}, J., {Kulkarni}, S.~R., \&
  {Backer}, D.~C. 1987, \nat, 328, 399, \dodoi{10.1038/328399a0}

\bibitem[{{Manchester} {et~al.}(2005){Manchester}, {Hobbs}, {Teoh}, \&
  {Hobbs}}]{2005AJ....129.1993M}
{Manchester}, R.~N., {Hobbs}, G.~B., {Teoh}, A., \& {Hobbs}, M. 2005, \aj, 129,
  1993, \dodoi{10.1086/428488}

\bibitem[{{McCarver} {et~al.}(2023){McCarver}, {Maccarone}, {Ransom}, {Clarke},
  {Giacintucci}, {Peters}, {Polisensky}, {Nyland}, {Gautam}, {Freire}, \&
  {Rangelov}}]{2023arXiv231211694M}
{McCarver}, A.~V., {Maccarone}, T.~J., {Ransom}, S.~M., {et~al.} 2023, arXiv
  e-prints, arXiv:2312.11694, \dodoi{10.48550/arXiv.2312.11694}

\bibitem[{{Nan} {et~al.}(2011){Nan}, {Li}, {Jin}, {Wang}, {Zhu}, {Zhu},
  {Zhang}, {Yue}, \& {Qian}}]{2011IJMPD..20..989N}
{Nan}, R., {Li}, D., {Jin}, C., {et~al.} 2011, International Journal of Modern
  Physics D, 20, 989, \dodoi{10.1142/S0218271811019335}

\bibitem[{{Nice} {et~al.}(2015){Nice}, {Demorest}, {Stairs}, {Manchester},
  {Taylor}, {Peters}, {Weisberg}, {Irwin}, {Wex}, \&
  {Huang}}]{2015ascl.soft09002N}
{Nice}, D., {Demorest}, P., {Stairs}, I., {et~al.} 2015, {Tempo: Pulsar timing
  data analysis}, Astrophysics Source Code Library, record ascl:1509.002.
\newblock \doeprint{1509.002}

\bibitem[{{Pan} {et~al.}(2020){Pan}, {Ransom}, {Lorimer}, {Fiore}, {Qian},
  {Wang}, {Stappers}, {Hobbs}, {Zhu}, {Yue}, {Wang}, {Lu}, {Liu}, {Peng},
  {Zhang}, \& {Li}}]{2020ApJ...892L...6P}
{Pan}, Z., {Ransom}, S.~M., {Lorimer}, D.~R., {et~al.} 2020, \apjl, 892, L6,
  \dodoi{10.3847/2041-8213/ab799d}

\bibitem[{{Pan} {et~al.}(2021{\natexlab{a}}){Pan}, {Ma}, {Qian}, {Wang}, {Yan},
  {Luo}, {Ransom}, {Lorimer}, \& {Jiang}}]{2021RAA....21..143P}
{Pan}, Z., {Ma}, X.-Y., {Qian}, L., {et~al.} 2021{\natexlab{a}}, Research in
  Astronomy and Astrophysics, 21, 143, \dodoi{10.1088/1674-4527/21/6/143}

\bibitem[{{Pan} {et~al.}(2021{\natexlab{b}}){Pan}, {Qian}, {Ma}, {Liu}, {Wang},
  {Luo}, {Yan}, {Ransom}, {Lorimer}, {Li}, \& {Jiang}}]{2021ApJ...915L..28P}
{Pan}, Z., {Qian}, L., {Ma}, X., {et~al.} 2021{\natexlab{b}}, \apjl, 915, L28,
  \dodoi{10.3847/2041-8213/ac0bbd}

\bibitem[{{Pan} {et~al.}(2023){Pan}, {Lu}, {Jiang}, {Han}, {Chen}, {Han},
  {Liu}, {Qian}, {Xu}, {Zhang}, {Luo}, {Yan}, {Yang}, {Zhou}, {Wang}, {Wang},
  {Li}, \& {Zhu}}]{2023Natur.620..961P}
{Pan}, Z., {Lu}, J.~G., {Jiang}, P., {et~al.} 2023, \nat, 620, 961,
  \dodoi{10.1038/s41586-023-06308-w}

\bibitem[{{Pooley} {et~al.}(2003){Pooley}, {Lewin}, {Anderson}, {Baumgardt},
  {Filippenko}, {Gaensler}, {Homer}, {Hut}, {Kaspi}, {Makino}, {Margon},
  {McMillan}, {Portegies Zwart}, {van der Klis}, \&
  {Verbunt}}]{2003ApJ...591L.131P}
{Pooley}, D., {Lewin}, W. H.~G., {Anderson}, S.~F., {et~al.} 2003, \apjl, 591,
  L131, \dodoi{10.1086/377074}

\bibitem[{{Qian} {et~al.}(2020){Qian}, {Yao}, {Sun}, {Xu}, {Pan}, \&
  {Jiang}}]{2020Innov...100053Q}
{Qian}, L., {Yao}, R., {Sun}, J., {et~al.} 2020, The Innovation, 1, 100053,
  \dodoi{10.1016/j.xinn.2020.100053}

\bibitem[{{Ransom}(2001)}]{2001PhDT.......123R}
{Ransom}, S.~M. 2001, PhD thesis, Harvard University, Massachusetts

\bibitem[{{Ransom}(2008)}]{2008AIPC..983..415R}
{Ransom}, S.~M. 2008, in American Institute of Physics Conference Series, Vol.
  983, 40 Years of Pulsars: Millisecond Pulsars, Magnetars and More, ed.
  C.~{Bassa}, Z.~{Wang}, A.~{Cumming}, \& V.~M. {Kaspi}, 415--423,
  \dodoi{10.1063/1.2900267}

\bibitem[{{Ransom} {et~al.}(2002){Ransom}, {Eikenberry}, \&
  {Middleditch}}]{2002AJ....124.1788R}
{Ransom}, S.~M., {Eikenberry}, S.~S., \& {Middleditch}, J. 2002, \aj, 124,
  1788, \dodoi{10.1086/342285}

\bibitem[{{Ransom} {et~al.}(2005){Ransom}, {Hessels}, {Stairs}, {Freire},
  {Camilo}, {Kaspi}, \& {Kaplan}}]{2005Sci...307..892R}
{Ransom}, S.~M., {Hessels}, J. W.~T., {Stairs}, I.~H., {et~al.} 2005, Science,
  307, 892, \dodoi{10.1126/science.1108632}

\bibitem[{{Ridolfi} {et~al.}(2021){Ridolfi}, {Gautam}, {Freire}, {Ransom},
  {Buchner}, {Possenti}, {Venkatraman Krishnan}, {Bailes}, {Kramer},
  {Stappers}, {Abbate}, {Barr}, {Burgay}, {Camilo}, {Corongiu}, {Jameson},
  {Padmanabh}, {Vleeschower}, {Champion}, {Chen}, {Geyer}, {Karastergiou},
  {Karuppusamy}, {Parthasarathy}, {Reardon}, {Serylak}, {Shannon}, \&
  {Spiewak}}]{2021MNRAS.504.1407R}
{Ridolfi}, A., {Gautam}, T., {Freire}, P.~C.~C., {et~al.} 2021, \mnras, 504,
  1407, \dodoi{10.1093/mnras/stab790}

\bibitem[{{Robinson} {et~al.}(1995){Robinson}, {Lyne}, {Manchester}, {Bailes},
  {D'Amico}, \& {Johnston}}]{1995MNRAS.274..547R}
{Robinson}, C., {Lyne}, A.~G., {Manchester}, R.~N., {et~al.} 1995, \mnras, 274,
  547, \dodoi{10.1093/mnras/274.2.547}

\bibitem[{{Schwab}(2021)}]{2021ApJ...906...53S}
{Schwab}, J. 2021, \apj, 906, 53, \dodoi{10.3847/1538-4357/abc87e}

\bibitem[{{Sigurdsson} {et~al.}(2003){Sigurdsson}, {Richer}, {Hansen},
  {Stairs}, \& {Thorsett}}]{2003Sci...301..193S}
{Sigurdsson}, S., {Richer}, H.~B., {Hansen}, B.~M., {Stairs}, I.~H., \&
  {Thorsett}, S.~E. 2003, Science, 301, 193, \dodoi{10.1126/science.1086326}

\bibitem[{{Tam} {et~al.}(2011){Tam}, {Kong}, {Hui}, {Cheng}, {Li}, \&
  {Lu}}]{2011ApJ...729...90T}
{Tam}, P.~H.~T., {Kong}, A.~K.~H., {Hui}, C.~Y., {et~al.} 2011, \apj, 729, 90,
  \dodoi{10.1088/0004-637X/729/2/90}

\bibitem[{{Turk} \& {Lorimer}(2013)}]{2013MNRAS.436.3720T}
{Turk}, P.~J., \& {Lorimer}, D.~R. 2013, \mnras, 436, 3720,
  \dodoi{10.1093/mnras/stt1850}

\bibitem[{{Verbunt} \& {Freire}(2014)}]{2014A&A...561A..11V}
{Verbunt}, F., \& {Freire}, P. C.~C. 2014, \aap, 561, A11,
  \dodoi{10.1051/0004-6361/201321177}

\bibitem[{{Wu} {et~al.}(2023){Wu}, {Pan}, {Qian}, {Ransom}, {Wang}, {Yan},
  {Luo}, {Zhang}, {Li}, {Yin}, {Li}, {Li}, {Dai}, {Li}, {Zhang}, {Liu}, \&
  {Pan}}]{2023arXiv231206067W}
{Wu}, Y., {Pan}, Z., {Qian}, L., {et~al.} 2023, arXiv e-prints,
  arXiv:2312.06067, \dodoi{10.48550/arXiv.2312.06067}

\bibitem[{{Yao} {et~al.}(2017){Yao}, {Manchester}, \&
  {Wang}}]{2017ApJ...835...29Y}
{Yao}, J.~M., {Manchester}, R.~N., \& {Wang}, N. 2017, \apj, 835, 29,
  \dodoi{10.3847/1538-4357/835/1/29}

\bibitem[{{Ye} {et~al.}(2019){Ye}, {Kremer}, {Chatterjee}, {Rodriguez}, \&
  {Rasio}}]{2019ApJ...877..122Y}
{Ye}, C.~S., {Kremer}, K., {Chatterjee}, S., {Rodriguez}, C.~L., \& {Rasio},
  F.~A. 2019, \apj, 877, 122, \dodoi{10.3847/1538-4357/ab1b21}

\bibitem[{{Ye} {et~al.}(2024){Ye}, {Kremer}, {Ransom}, \&
  {Rasio}}]{2024ApJ...961...98Y}
{Ye}, C.~S., {Kremer}, K., {Ransom}, S.~M., \& {Rasio}, F.~A. 2024, \apj, 961,
  98, \dodoi{10.3847/1538-4357/ad089a}

\bibitem[{{Yin} {et~al.}(2023){Yin}, {Zhang}, {Li}, {Li}, {Qian}, \&
  {Pan}}]{2023RAA....23e5012Y}
{Yin}, D.-J., {Zhang}, L.-Y., {Li}, B.-D., {et~al.} 2023, Research in Astronomy
  and Astrophysics, 23, 055012, \dodoi{10.1088/1674-4527/acc37e}

\bibitem[{{Zhang} {et~al.}(2022){Zhang}, {Ridolfi}, {Blumer}, {Freire},
  {Manchester}, {McLaughlin}, {Kremer}, {Cameron}, {Zhang}, {Behrend},
  {Burgay}, {Buchner}, {Champion}, {Chen}, {Dai}, {Feng}, {Fu}, {Guo}, {Hobbs},
  {Keane}, {Kramer}, {Levin}, {Li}, {Ni}, {Pan}, {Padmanabh}, {Possenti},
  {Ransom}, {Tsai}, {Venkatraman Krishnan}, {Wang}, {Zhang}, {Zhi}, {Zhang}, \&
  {Li}}]{2022ApJ...934L..21Z}
{Zhang}, L., {Ridolfi}, A., {Blumer}, H., {et~al.} 2022, \apjl, 934, L21,
  \dodoi{10.3847/2041-8213/ac81c3}

\bibitem[{{Zhao} \& {Heinke}(2022)}]{2022MNRAS.511.5964Z}
{Zhao}, J., \& {Heinke}, C.~O. 2022, \mnras, 511, 5964,
  \dodoi{10.1093/mnras/stac442}

\end{thebibliography}
\bibliographystyle{aasjournal}

\begin{longtable}{{lllllllllll}}
\caption{The 45 GCs in FAST sky.
The latest central escape velocities ($V_{\rm esc}$) of GCs are from the Baumgardt's list (The $4^{\rm th}$ Mar. 2023) version\footnote{\url{https://people.smp.uq.edu.au/HolgerBaumgardt/globular/parameter.html}}), while other physical parameters of GCs are the same as in \citet{2013MNRAS.436.3720T}. 
The $N_{\rm obs}$ is the number of observed known pulsars by the end of 2023.
The expected number of potential pulsars of $\hat{N}_{\rm psr, 1}$ and $\hat{N}_{\rm psr, 2}$ 
are from the Equation 13 ($\ln\hat{\lambda}=-1.1+1.5\lg\Gamma$) of \citet{2013MNRAS.436.3720T} and model no. 2 from the FAST sky only sample in this work ($\ln\hat{N}_{\rm psr, i}=   0.369 + 0.087\  V_{\rm esc, i}$ ), respectively.
The $D_{\rm Sun}$ is the distance from the Sun (\citealt{1996AJ....112.1487H}, 2010 edition).
The $L_{\rm min}$ ($L_{\rm min,i}$=$S_{\rm min,i}$$D^2_{\rm Sun,i}$, \citealt{2011ApJ...742...51B}) is from the list of flux density detection limits of GCs.
Note: Table 4 and Table 5 are also can be presented in an online available form.
}
\label{table:longtable_example} \\
\hline
ID &  Name & $D_{\rm Sun}$ & $L_{\rm min}$ & $\Gamma$  & $V_{\rm esc}$ & $N_{\rm obs}$ & $\hat{N}_{\rm psr, 1}$ & $R_1$ & $\hat{N}_{\rm psr, 2}$ & $R_2$ \\ & &  (kpc)  &  (mJy kpc$^2$)&   & (km s$^{-1}$) &  &   &  ($N_{\rm obs}$/$\hat{N}_{\rm psr, 1}$)   &  & ($N_{\rm obs}$/$\hat{N}_{\rm psr, 2}$)   \\  
\hline 
\endfirsthead
\multicolumn{9}{c}%
{{\bfseries \tablename\ \thetable{} -- continued from previous page}} \\
\hline 
ID &  Name & $D_{\rm Sun}$ & $L_{\rm min}$ & $\Gamma$  & $V_{\rm esc}$ & $N_{\rm obs}$ & $\hat{N}_{\rm psr, 1}$ & $R_1$ & $\hat{N}_{\rm psr, 2}$ & $R_2$ \\ & &  (kpc)  &  (mJy kpc$^2$)&   & (km s$^{-1}$) &  &   &  ($N_{\rm obs}$/$\hat{N}_{\rm psr, 1}$)   &  & ($N_{\rm obs}$/$\hat{N}_{\rm psr, 2}$)    \\
\hline  
\endhead

\hline \multicolumn{9}{r}{{Continued on next page}} \\ \hline
\endfoot

\hline \hline
\endlastfoot

NGC 7078  & M 15  & 10.4   & 2.2714  & 4510    & 48.9 & 12 & 80  & 0.15  & 102 & 0.12 \\
NGC 7089  & M 2   & 11.5   & 0.7935  & 518     & 43.6 & 6  & 20  & 0.3   & 64  & 0.09 \\
NGC 6517  &       & 10.6   & 0.3371  & 338     & 37.3 & 17 & 15  & 1.13  & 37  & 0.46 \\
NGC 6341  & M 92  & 8.3    & 0.3513  & 270     & 36.4 & 1  & 13  & 0.08  & 34  & 0.03 \\
NGC 6402  & M 14  & 9.3    & 4.7397  & 124     & 35.6 & 5  & 8   & 0.63  & 32  & 0.16 \\
NGC 6205  & M 13  & 7.1    & 1.3611  & 68.9    & 32.4 & 6  & 5   & 1.2   & 24  & 0.25 \\
NGC 5272  & M 3   & 10.2   & 2.1848  & 194     & 32   & 6  & 10  & 0.6   & 23  & 0.26 \\
NGC 5904  & M 5   & 7.5    & 1.4063  & 164     & 30.4 & 7  & 9   & 0.78  & 20  & 0.35 \\
NGC 5024  & M 53  & 17.9   & 6.0878  & 35.4    & 25.9 & 5  & 3   & 1.67  & 14  & 0.36 \\
NGC 6760  &       & 7.4    & 2.0261  & 56.9    & 25.6 & 2  & 5   & 0.4   & 13  & 0.15 \\
NGC 6229  &       & 30.5   &         & 47.6    & 25.3 &    & 4   & 0     & 13  & 0    \\
NGC 5634  &       & 25.2   &         & 20.2    & 24.2 &    & 2   & 0     & 12  & 0    \\
NGC 6254  & M 10  & 4.4    & 1.2894  & 31.4    & 23.9 & 2  & 3   & 0.67  & 12  & 0.17 \\
NGC 6539  &       & 7.8    & 2.8595  & 42.1    & 22.5 & 1  & 4   & 0.25  & 10  & 0.1  \\
NGC 6779  & M 56  & 9.4    & 2.209   & 27.7    & 21.9 &    & 3   & 0     & 10  & 0    \\
NGC 6934  &       & 15.6   & 6.8141  & 29.9    & 21.4 &    & 3   & 0     & 9   & 0    \\
NGC 6749  &       & 7.9    & 1.9971  & 38.5    & 20.5 & 2  & 4   & 0.5   & 9   & 0.22 \\
NGC 6712  &       & 6.9    & 0.319   & 30.8    & 19.2 & 1  & 3   & 0.33  & 8   & 0.13 \\
Pal 2     &       & 27.2   & 25.1546 & 929     & 19.2 &    & 29  & 0     & 8   & 0    \\
NGC 2419  &       & 82.6   &         & 2.8     & 18.2 &    & 1   & 0     & 7   & 0    \\
NGC 6218  & M 12  & 4.8    & 1.2626  & 13      & 17.6 & 2  & 2   & 1     & 7   & 0.29 \\
NGC 6171  & M 107 & 6.4    & 2.2446  & 6.77    & 15.7 &    & 1   & 0     & 6   & 0    \\
NGC 7006  &       & 41.2   & 32.2514 & 9.4     & 15.7 &    & 1   & 0     & 6   & 0    \\
Pal 10    &       & 5.9    & 0.8006  & 59      & 14.6 &    & 5   & 0     & 5   & 0    \\
IC 1276   & Pal 7 & 5.4    &         & 7.97    & 14.3 &    & 1   & 0     & 5   & 0    \\
NGC 4147  &       & 19.3   & 7.0773  & 16.6    & 14.2 &    & 2   & 0     & 5   & 0    \\
NGC 6981  & M 72  & 17     & 1.734   & 4.69    & 12.4 &    & 1   & 0     & 4   & 0    \\
NGC 6426  &       & 20.6   & 10.609  & 1.58    & 10.3 &    & 0   & -     & 4   & 0    \\
NGC 6535  &       & 6.8    & 1.8958  & 0.388   & 10.2 &    & 0   & -     & 4   & 0    \\
NGC 6838  & M 71  & 4      & 0.304   & 2.05    & 10.2 & 5  & 1   & 5     & 4   & 1.25 \\
NGC 6366  &       & 3.5    &         & 5.14    & 9.1  &    & 1   & 0     & 3   & 0    \\
IC 1257   &       & 25     &         &         & 6.7  &    & -   & -     & 3   & 0    \\
NGC 5466  &       & 16     & 5.632   & 0.239   & 6.5  &    & 0   & -     & 3   & 0    \\
NGC 5053  &       & 17.4   & 6.0552  & 0.105   & 6    &    & 0   & -     & 2   & 0    \\
Pal 15    &       & 45.1   & 77.2924 & 0.0222  & 4.3  &    & 0   & -     & 2   & 0    \\
Pal 11    &       & 13.4   &         & 20.8    & 4    &    & 2   & 0     & 2   & 0    \\
Pal 4     &       & 108.7  &         & 0.0189  & 2.6  &    & 0   & -     & 2   & 0    \\
Pal 3     &       & 92.5   &         & 0.0409  & 2.4  &    & 0   & -     & 2   & 0    \\
Pal 14    & AvdB  & 76.5   &         & 0.00186 & 2.3  &    & 0   & -     & 2   & 0    \\
Pal 5     &       & 23.2   & 17.2237 & 0.00212 & 2.1  &    & 0   & -     & 2   & 0    \\
Pal 13    &       & 26     & 14.196  & 0.00109 & 1.8  &    & 0   & -     & 2   & 0    \\
Whiting 1 &       & 30.1   &         &         & 1.1  &    & -   & -     & 2   & 0    \\
GLIMPSE01 &       & 4.2    &         &         &      & 2  & -   & -     & -   & -    \\
Ko 1      &       & 48.3   &         &         &      &    & -   & -     & -   & -    \\
Ko 2      &       & 34.7   &         &         &      &    & -   & -     & -   & -    \\
\end{longtable}

\begin{longtable}{{lllllllllll}}
\caption{The 112 GCs outside of FAST aky}
\label{table:longtable_example_1} \\
\hline
ID &  Name & $D_{\rm Sun}$ & $L_{\rm min}$ & $\Gamma$  & $V_{\rm esc}$ & $N_{\rm obs}$ & $\hat{N}_{\rm psr, 1}$ & $R_1$ & $\hat{N}_{\rm psr, 2}$ & $R_2$ \\ & &  (kpc)  &  (mJy kpc$^2$)&   & (km s$^{-1}$) &  &   &  ($N_{\rm obs}$/$\hat{N}_{\rm psr, 1}$)   &  & ($N_{\rm obs}$/$\hat{N}_{\rm psr, 2}$)    \\
\hline 
\endfirsthead
\multicolumn{9}{c}%
{{\bfseries \tablename\ \thetable{} -- continued from previous page}} \\
\hline 
ID &  Name & $D_{\rm Sun}$ & $L_{\rm min}$ & $\Gamma$  & $V_{\rm esc}$ & $N_{\rm obs}$ & $\hat{N}_{\rm psr, 1}$ & $R_1$ & $\hat{N}_{\rm psr, 2}$ & $R_2$ \\ & &  (kpc)  &  (mJy kpc$^2$)&   & (km s$^{-1}$) &  &   &  ($N_{\rm obs}$/$\hat{N}_{\rm psr, 1}$)   &  & ($N_{\rm obs}$/$\hat{N}_{\rm psr, 2}$)    \\
\hline  
\endhead

\hline \multicolumn{9}{r}{{Continued on next page}} \\ \hline
\endfoot

\hline \hline
\endlastfoot

 Liller 1  &             & 8.2   & 0.6657  & 391     & 87.9 &    & 16  & 0    & 3030 & 0    \\
NGC 6441  &             & 11.6  & 1.6685  & 2300    & 69.4 & 9  & 52  & 0.17 & 606  & 0.01 \\
NGC 6715  & M 54        & 26.5  &         & 2520    & 68.9 &    & 55  & 0    & 580  & 0    \\
NGC 5139  & $\omega$-Cen   & 5.2   & 1.306   & 90.4    & 62.2 & 18 & 6   & 3    & 324  & 0.06 \\
NGC 6388  &             & 9.9   & 5.3709  & 899     & 60   &    & 28  & 0    & 267  & 0    \\
NGC 6266  & M 62        & 6.8   & 1.0219  & 1670    & 59.3 & 10 & 42  & 0.24 & 252  & 0.04 \\
Terzan 5  & Terzan 11   & 6.9   & 0.5523  & 6800    & 57.5 & 49 & 104 & 0.47 & 215  & 0.23 \\
NGC 6440  &             & 8.5   & 0.7081  & 1400    & 55.6 & 8  & 37  & 0.22 & 182  & 0.04 \\
NGC 2808  &             & 9.6   & 5.0504  & 923     & 53.1 &    & 28  & 0    & 147  & 0    \\
NGC 6864  & M 75        & 20.9  &         & 307     & 48.4 &    & 14  & 0    & 97   & 0    \\
NGC 104   & 47 Tuc      & 4.5   & 3.3939  & 1000    & 47.4 & 35 & 30  & 1.17 & 89   & 0.39 \\
NGC 6139  &             & 10.1  & 7.9058  & 307     & 46.4 &    & 14  & 0    & 82   & 0    \\
NGC 5824  &             & 32.1  &         & 984     & 44.8 &    & 30  & 0    & 71   & 0    \\
NGC 6273  & M 19        & 8.8   & 4.2437  & 200     & 44.8 &    & 11  & 0    & 71   & 0    \\
NGC 6093  & M 80        & 10    & 0.57    & 532     & 44.6 & 1  & 20  & 0.05 & 70   & 0.01 \\
NGC 5286  &             & 11.7  & 7.5016  & 458     & 41.8 &    & 18  & 0    & 55   & 0    \\
NGC 1851  &             & 12.1  & 4.4069  & 1530    & 41.1 & 15 & 40  & 0.38 & 52   & 0.29 \\
NGC 6626  & M 28        & 5.5   & 0.124   & 648     & 41.1 & 14 & 23  & 0.61 & 52   & 0.27 \\
Terzan 1  & HP 2        & 6.7   & 3.479   & 0.292   & 37.9 & 7  & 0   & -    & 39   & 0.18 \\
NGC 6541  &             & 7.5   & 0.54    & 386     & 37.3 &    & 16  & 0    & 37   & 0    \\
NGC 6656  & M 22        & 3.2   & 0.0584  & 77.5    & 35.5 & 4  & 6   & 0.67 & 32   & 0.13 \\
Terzan 10 &             & 5.8   &         & 4430    & 35.2 &    & 79  & 0    & 31   & 0    \\
NGC 362   &             & 8.6   &         & 735     & 34   & 12 & 25  & 0.48 & 28   & 0.43 \\
NGC 5694  &             & 35    &         & 191     & 33.6 &    & 10  & 0    & 27   & 0    \\
NGC 6553  &             & 6     &         & 69      & 32.8 &    & 5   & 0    & 25   & 0    \\
NGC 6522  &             & 7.7   & 3.2491  & 363     & 32.4 & 6  & 15  & 0.4  & 24   & 0.25 \\
NGC 6333  & M 9         & 7.9   & 3.4201  & 131     & 32.4 &    & 8   & 0    & 24   & 0    \\
NGC 6380  & Ton 1       & 10.9  &         & 116     & 31.8 &    & 7   & 0    & 23   & 0    \\
NGC 6752  &             & 4     & 0.504   & 401     & 31.3 & 9  & 17  & 0.53 & 22   & 0.41 \\
NGC 6356  &             & 15.1  &         & 88.1    & 31.1 &    & 6   & 0    & 22   & 0    \\
NGC 5986  &             & 10.4  & 0.4002  & 61.9    & 30.6 & 1  & 5   & 0.2  & 21   & 0.05 \\
NGC 6316  &             & 10.4  &         & 77      & 30.3 & 1  & 6   & 0.17 & 20   & 0.05 \\
Terzan 9  &             & 7.1   &         & 1.71    & 29.1 &    & 0   & -    & 18   & 0    \\
NGC 6569  &             & 10.9  &         & 53.6    & 28.4 &    & 4   & 0    & 17   & 0    \\
NGC 6624  &             & 7.9   & 0.9986  & 1150    & 28   & 12 & 33  & 0.36 & 17   & 0.71 \\
NGC 6453  &             & 11.6  & 10.8186 & 371     & 27.9 &    & 16  & 0    & 16   & 0    \\
NGC 6284  &             & 15.3  & 12.8281 & 666     & 27.6 &    & 23  & 0    & 16   & 0    \\
NGC 1904  & M 79        & 12.9  &         & 116     & 27.6 &    & 7   & 0    & 16   & 0    \\
NGC 6293  &             & 9.5   & 4.9457  & 847     & 27.4 &    & 27  & 0    & 16   & 0    \\
NGC 6287  &             & 9.4   & 6.8479  & 36.3    & 26.8 &    & 3   & 0    & 15   & 0    \\
NGC 6638  &             & 9.4   &         & 137     & 26.7 &    & 8   & 0    & 15   & 0    \\
Terzan 6  & HP 5        & 6.8   & 0.3376  & 2470    & 26.4 &    & 54  & 0    & 14   & 0    \\
NGC 6681  & M 70        & 9     & 6.2775  & 1040    & 25.3 &    & 31  & 0    & 13   & 0    \\
NGC 5927  &             & 7.7   & 3.2491  & 68.2    & 25.2 &    & 5   & 0    & 13   & 0    \\
NGC 6544  &             & 3     & 0.3492  & 111     & 25.2 & 3  & 7   & 0.43 & 13   & 0.23 \\
NGC 5946  &             & 10.6  & 6.1573  & 134     & 24.3 &    & 8   & 0    & 12   & 0    \\
NGC 6637  & M 69        & 8.8   &         & 89.9    & 24.3 &    & 6   & 0    & 12   & 0    \\
NGC 6401  &             & 10.6  & 8.7079  & 44      & 23.6 &    & 4   & 0    & 11   & 0    \\
Terzan 4  & HP 4        & 7.2   &         &         & 23.1 &    & -   & -    & 11   & 0    \\
NGC 6304  &             & 5.9   & 1.9076  & 123     & 22.8 &    & 8   & 0    & 11   & 0    \\
NGC 4833  &             & 6.6   & 3.3759  & 25      & 22.2 &    & 3   & 0    & 10   & 0    \\
FSR 1735  &             & 9.8   &         &         & 22.2 &    & -   & -    & 10   & 0    \\
NGC 6355  &             & 9.2   & 6.5596  & 99.2    & 21.6 &    & 7   & 0    & 9    & 0    \\
NGC 6397  &             & 2.3   & 0.1666  & 84.1    & 21.5 & 2  & 6   & 0.33 & 9    & 0.22 \\
NGC 6723  &             & 8.7   & 4.1478  & 11.4    & 21.5 &    & 2   & 0    & 9    & 0    \\
NGC 1261  &             & 16.3  & 10.3088 & 15.4    & 21.4 &    & 2   & 0    & 9    & 0    \\
NGC 7099  & M 30        & 8.1   & 0.7086  & 324     & 21   & 2  & 14  & 0.14 & 9    & 0.22 \\
NGC 6325  &             & 7.8   & 3.334   & 118     & 20.6 &    & 7   & 0    & 9    & 0    \\
HP 1      & BH 229      & 8.2   &         & 0.662   & 20.3 &    & 0   & -    & 8    & 0    \\
NGC 6642  &             & 8.1   & 3.5954  & 97.8    & 20   &    & 7   & 0    & 8    & 0    \\
NGC 6256  &             & 10.3  &         & 169     & 19.8 &    & 9   & 0    & 8    & 0    \\
Pal 6     &             & 5.8   & 0.2254  & 15.5    & 19.7 &    & 2   & 0    & 8    & 0    \\
NGC 6652  &             & 10    & 7.75    & 700     & 19.1 & 2  & 24  & 0.08 & 8    & 0.25 \\
NGC 6528  &             & 7.9   & 3.4201  & 278     & 19   &    & 13  & 0    & 8    & 0    \\
NGC 6121  & M 4         & 2.2   & 0.636   & 26.9    & 18.8 & 1  & 3   & 0.33 & 7    & 0.14 \\
NGC 6342  &             & 8.5   & 19.5075 & 44.8    & 18.4 & 2  & 4   & 0.5  & 7    & 0.29 \\
NGC 6809  & M 55        & 5.4   & 2.2599  & 3.23    & 18.1 &    & 1   & 0    & 7    & 0    \\
Terzan 2  & HP 3        & 7.5   &         & 22.1    & 17.4 &    & 3   & 0    & 7    & 0    \\
NGC 3201  &             & 4.9   & 1.3157  & 7.17    & 17.3 &    & 1   & 0    & 7    & 0    \\
NGC 6584  &             & 13.5  & 9.9873  & 11.8    & 16.7 &    & 2   & 0    & 6    & 0    \\
NGC 6558  &             & 7.4   & 3.4663  & 105     & 16.7 &    & 7   & 0    & 6    & 0    \\
UKS 1     &             & 7.8   & 3.334   & 100     & 15.7 &    & 7   & 0    & 6    & 0    \\
NGC 4372  &             & 5.8   & 1.8435  & 0.233   & 15.5 &    & 0   & -    & 6    & 0    \\
NGC 6235  &             & 11.5  & 7.2473  & 5.75    & 15.5 &    & 1   & 0    & 6    & 0    \\
Djorg 2   & ESO456-SC38 & 6.3   &         & 46.4    & 15.4 &    & 4   & 0    & 6    & 0    \\
NGC 2298  &             & 10.8  & 0.4782  & 4.31    & 15.3 &    & 1   & 0    & 5    & 0    \\
NGC 6540  & Djorg 3     & 5.3   & 2.177   & 263     & 15.1 &    & 13  & 0    & 5    & 0    \\
NGC 6144  &             & 8.9   & 4.3407  & 3.14    & 15   &    & 1   & 0    & 5    & 0    \\
Djorg 1   &             & 13.7  &         & 9.04    & 14.8 &    & 1   & 0    & 5    & 0    \\
NGC 4590  & M 68        & 10.3  & 5.8137  & 5.82    & 14.8 &    & 1   & 0    & 5    & 0    \\
NGC 6352  &             & 5.6   &         & 6.74    & 13.2 &    & 1   & 0    & 5    & 0    \\
NGC 6362  &             & 7.6   &         & 4.56    & 13.1 &    & 1   & 0    & 5    & 0    \\
Terzan 12 &             & 4.8   &         & 35.8    & 13.1 &    & 3   & 0    & 5    & 0    \\
Lynga 7   & BH184       & 8     &         &         & 12.8 &    & -   & -    & 4    & 0    \\
NGC 5897  &             & 12.5  & 0.8594  & 0.851   & 12.5 &    & 0   & -    & 4    & 0    \\
Pal 8     &             & 12.8  &         & 4.22    & 11.7 &    & 1   & 0    & 4    & 0    \\
Ton 2     & Pismis 26   & 8.2   &         & 4.29    & 11.6 &    & 1   & 0    & 4    & 0    \\
NGC 6717  & Pal 9       & 7.1   & 2.7625  & 39.8    & 11.6 &    & 4   & 0    & 4    & 0    \\
NGC 6101  &             & 15.4  &         & 0.974   & 11.5 &    & 0   & -    & 4    & 0    \\
2MS-GC01  & 2MASS-GC01  & 3.6   &         &         & 11.2 &    & -   & -    & 4    & 0    \\
NGC 288   &             & 8.9   & 0.5149  & 0.766   & 10.9 &    & 0   & -    & 4    & 0    \\
NGC 6496  &             & 11.3  & 6.9974  & 0.657   & 10.8 &    & 0   & -    & 4    & 0    \\
IC 4499   &             & 18.8  &         & 0.797   & 10.5 &    & 0   & -    & 4    & 0    \\
BH 261    & AL 3        & 6.5   &         &         & 9.1  &    & -   & -    & 3    & 0    \\
2MS-GC02  & 2MASS-GC02  & 4.9   &         &         & 8.4  &    & -   & -    & 3    & 0    \\
Terzan 3  &             & 8.2   &         & 0.89    & 7.6  &    & 0   & -    & 3    & 0    \\
ESO-SC06  & ESO280-SC06 & 21.4  &         &         & 6.8  &    & -   & -    & 3    & 0    \\
Terzan 8  &             & 26.3  &         & 0.0793  & 6    &    & 0   & -    & 2    & 0    \\
Rup 106   &             & 21.2  &         & 0.359   & 5.6  &    & 0   & -    & 2    & 0    \\
1636-283  & ESO452-SC11 & 8.3   &         & 1.72    & 5.5  &    & 0   & -    & 2    & 0    \\
Terzan 7  &             & 22.8  &         & 1.59    & 4.6  &    & 0   & -    & 2    & 0    \\
Arp 2     &             & 28.6  &         & 0.00518 & 4.5  &    & 0   & -    & 2    & 0    \\
NGC 7492  &             & 26.3  &         & 0.192   & 4.4  &    & 0   & -    & 2    & 0    \\
Pyxis     &             & 39.4  &         &         & 3.6  &    & -   & -    & 2    & 0    \\
AM 1      & E 1         & 123.3 &         & 0.00419 & 3.3  &    & 0   & -    & 2    & 0    \\
Pal 12    &             & 19    & 2.2382  & 0.397   & 2.5  &    & 0   & -    & 2    & 0    \\
E 3       &             & 8.1   &         & 0.0759  & 2.4  &    & 0   & -    & 2    & 0    \\
Eridanus  &             & 90.1  &         & 0.0339  & 2.2  &    & 0   & -    & 2    & 0    \\
Pal 1     &             & 11.1  &         & 0.895   & 1.9  &    & 0   & -    & 2    & 0    \\
AM 4      &             & 32.2  &         & 0.0026  & 0.9  &    & 0   & -    & 2    & 0    \\
BH 176    &             & 18.9  &         & 0.146   &      &    & 0   & -    & -    & -    \\
GLIMPSE02 &             & 5.5   &         &         &      &    & -   & -    & -    & -    \\

\end{longtable}

\end{document}